\documentclass[10pt, aps, prd, onecolumn, amsmath, amssymb, floatfix, superscriptaddress, nofootinbib]{revtex4-2}

\usepackage[utf8]{inputenc}
\usepackage[T1]{fontenc}
\usepackage{graphicx}
\usepackage{bm}
\usepackage{color}
\usepackage[normalem]{ulem}
\usepackage{booktabs}
\usepackage{multirow}
\usepackage{aas_macros} 
\usepackage{dsfont}
\usepackage{hyperref}
\hypersetup{colorlinks=true, linkcolor=blue, citecolor=blue, urlcolor=cyan}
\graphicspath{{figs/}}

\newcommand{\Msun}{\ensuremath{M_\odot}}
\newcommand{\fmc}{\ensuremath{\mathrm{fm}^{-3}}}
\newcommand{\cs}{\ensuremath{c_s^2}}

\begin{document}

\title{Set Transformer inference of the neutron star equation of state \\ from stellar observations}

\author{Márcio Ferreira}
\email{marcio.ferreira@uc.pt}
\affiliation{CFisUC, 
	Department of Physics, University of Coimbra, P-3004 - 516  Coimbra, Portugal}

\author{Valéria Carvalho}
\email{val.mar.dinis@uc.pt}
\affiliation{CFisUC, 
	Department of Physics, University of Coimbra, P-3004 - 516  Coimbra, Portugal}
\affiliation{Nicolaus Copernicus Astronomical Center, Polish Academy of Sciences, Bartycka 18, 00-716, Warsaw, Poland}

\author{Micha{\l} Bejger}
\email{bejger@camk.edu.pl}
\affiliation{Nicolaus Copernicus Astronomical Center, Polish Academy of Sciences, Bartycka 18, 00-716, Warsaw, Poland}
\affiliation{INFN Sezione di Ferrara, Via Saragat 1, 44122 Ferrara, Italy}

\author{Constança Providência}
\email{cp@uc.pt}
\affiliation{CFisUC, 
	Department of Physics, University of Coimbra, P-3004 - 516  Coimbra, Portugal}
\date{\today}

\begin{abstract}
We develop a permutation-invariant Set Transformer to reconstruct the
equation of state (EoS) of dense matter from variable-size, unordered sets
of neutron star (NS) observations. The model takes stellar masses together
with radii, tidal deformabilities, or both, and predicts either the
pressure $P(n)$ or the sound speed $c_s^2(n)$ on a fixed density
grid, along with density-dependent uncertainties. Nothing in the
architecture prescribes which star informs which density: self-attention
couples all observations nonlinearly, and each density point reads the full
set through its own learnable query, so the star-to-density mapping is
learned from the data. Trained on independent piecewise-polytropic and
Gaussian-process EoS ensembles, the model provides well-calibrated
predictions whose uncertainty increases in density regions that stable
stars cannot probe. Reconstruction errors decrease with the number of
observations, while tidal deformability generally improves accuracy at
a fixed observation count, even when it carries its own measurement noise.
Sensitivity analysis reveals a density-local mapping: in the pressure
models, predictions at density $n$ depend most strongly on stars whose
central densities are near $n$. We also show that the sensitivity of the model to the inferred stellar compactness provides information on the minimum central density. These results demonstrate that set-based
neural inference, in which the star-to-density mapping is learned rather
than assumed, can extract physically interpretable EoS information with
calibrated uncertainties. 
\end{abstract}

\maketitle

%======================================================================
\section{Introduction}
\label{sec:intro}
%======================================================================
The equation of state (EoS) of cold and dense matter determines the structure and
observable properties of neutron stars (NSs) \cite{Haensel2007NeutronStars1,glendenning2012compact,AnderssonN2019}. In particular, the relation between
pressure $P$ and energy density ${\cal E}$ (which differs from the rest-mass
baryon density $n$ by the internal and binding energy of the matter) determines the stellar mass--radius $M(R)$
relation through the Tolman-Oppenheimer-Volkoff (TOV) equations
\cite{PhysRev.55.364,PhysRev.55.374}. The same stellar structure calculation
also gives the tidal deformability $\Lambda=\tfrac{2}{3}k_2\,C^{-5}$, with
$k_2$ the quadrupolar Love number and the stellar compactness $C=GM/(Rc^2)$
\cite{1911spge.book.....L,2008PhRvD..77b1502F,2017PhRvD..95h3014V}. Consequently, measurements of
stellar masses, radii, and tidal deformabilities constrain the EoS only
through this forward map, making the reconstruction of $P({\cal E})$ from a finite
set of stars an inverse problem. At zero temperature ${\cal E}$ and $n$ 
are uniquely related, so reconstructing $P({\cal E})$ and $P(n)$ are
equivalent problems; we work with $P(n)$ throughout.

Current observations already probe complementary density regimes. Massive
pulsars, including PSR~J0348$-$0432, PSR~J0740$+$6620 and PSR~J1614-2230 \cite{Antoniadis2013,Fonseca:2021wxt,mauviard2026nicer}, constrain the high-density EoS through
the approximately $2\,M_\odot$ maximum-mass bound. NICER radius measurements
\cite{Riley_2019,Miller19,Riley2021,Miller2021,choudhury2024nicer,salmi2024nicer,mauviard2025nicer} and tidal deformability
constraints from GW170817 \cite{Abbott:2018wiz} provide additional information
at intermediate densities. Upcoming X-ray and radio facilities
\cite{eXTP,STROBE-X,SKA} will deliver larger and more precise samples, making
methods that can combine observations from multiple stars increasingly
important.

Bayesian inference remains the standard approach for reconstructing the EoS.
It can provide a full posterior while allowing for flexible representations,
including piecewise-polytropic, sound speed, and Gaussian-process models
\cite{Landry:2018prl,Essick2019,Landry:2020vaw}. However, such analyses are
computationally expensive and generally must be repeated when the observation
set changes. Machine-learning methods offer a complementary route: neural
networks and other surrogate models have been used to infer EoS properties
from NS observables \cite{Ferreira:2022nwh,Fujimoto_2021,Fujimoto:2019hxv,Fujimoto_2018,
Soma:2022qnv,Carvalho:2023ele,Carvalho:2024kgf,Soma:2022vbb,baker2026physicsinformedbayesianneural}, while normalizing flows and
conditional variational autoencoders have been used for probabilistic EoS
reconstruction \cite{Brandes:2024vhw,Ferreira:2024rnf,Carvalho:2024zyb,carvalho2025neural,Carvalho:2026nbl}.
Recent work has also addressed uncertainty quantification in
machine learning EoS inference \cite{Fujimoto:2024cyv}. Once trained, these
models can evaluate new observation sets at negligible additional cost.

A collection of NS observations is naturally represented as a set: its
elements have no intrinsic ordering, and its size can vary as observations
accumulate. These properties are not respected by conventional fixed-length
architectures unless padding or ad hoc ordering is introduced. The Set
Transformer \cite{lee2019set} provides a natural alternative, combining
permutation invariance with attention-based interactions between set elements
and supporting variable-size inputs. Its attention mechanism allows the model to capture interactions between individual stellar observations while forming a global representation of the entire observation set. Transformer based architectures have also recently been explored for gravitational wave parameter estimation \cite{kofler2026flexible,papalini2025can,Leyde:2026hvm}.

In this work, we use a Set Transformer to reconstruct two dense matter
functionals: the pressure $P(n)$ and the sound speed squared
$c_s^2(n)$ as a function of baryonic density. For each target, the model predicts the full profile on a fixed
density grid, together with a density-dependent uncertainty using a
heteroscedastic Gaussian output head (\emph{heteroscedastic} meaning that the predicted variance is not fixed but varies from point to point, in contrast to a \emph{homoscedastic}  model that assumes a single, constant noise level everywhere). We compare mass--radius and mass--tidal-deformability observations under several noise levels and test the
method on independent piecewise-polytropic and Gaussian-process (GP) EoS ensembles.

Beyond reconstruction accuracy, we examine calibration and the physical
content of the learned mapping through input-output sensitivities. We find a
density-local structure: predictions at density $n$ are most sensitive to
stars whose central densities are close to $n$. We further show that
compactness provides a more universal coordinate than mass for relating a
stellar observation to the density it probes. These results establish a
permutation-invariant framework for fast, uncertainty-aware, and physically
interpretable EoS inference from sets of NS observations.

The paper is organized as follows. Section~\ref{sec:model} describes the
architecture, training procedure, model variants, and EoS ensembles.
Section~\ref{sec:results} presents the results for the pressure target
(Sec.~\ref{sec:pressure}) and the sound speed target
(Sec.~\ref{sec:soundspeed}). Section~\ref{sec:conclusions} summarizes our
conclusions.

%======================================================================
\section{Model and data}
\label{sec:model}
%======================================================================

\subsection{Set Transformer architecture}
\label{sec:arch}
We use a Set Transformer neural network as an inference framework
for the EoS of NS matter. The model takes an unordered set of $N\in\{5,\dots,35\}$ observations and
returns the EoS at $20$ baryon densities, evenly spaced between 
$n_1 = 0.10\ \fmc$ and $n_{20} = 1.10\ \fmc$, so that the grid spacing is
$\Delta n \simeq 0.053\ \fmc$.
When both observables, $R(M)$ and $\Lambda(M)$, are used, each set
contains $2N$ elements: $N$ radius measurements and $N$
tidal-deformability measurements, drawn independently from the same EoS
and thus generally referring to different stars. The network reads every
element in the same format, the four-component feature vector
$x_i = (M_i,\, R_i,\, \Lambda_i,\, \mathds{1}_\Lambda)$, so each element
fills only the slots corresponding to its own measurement: a radius
measurement is stored as $(M_i,\, R_i,\, 0,\, 0)$ and a
tidal-deformability measurement as $(M_i,\, 0,\, \Lambda_i,\, 1)$, the
unused entry being set to zero. 
The binary flag $\mathds{1}_\Lambda$ ---
$1$ for a $\Lambda(M)$ element and $0$ for an $R(M)$ element ---  explicitly identifies the measurement type.
In particular, no element ever contains both an $R$ and a
$\Lambda$ measurement: the model must infer the connection between the two
observables from their joint dependence on the underlying EoS.
No positional encoding is applied, so the network is permutation-invariant.

\begin{figure}[!ht]
    \centering
    \includegraphics[width=0.5\linewidth]{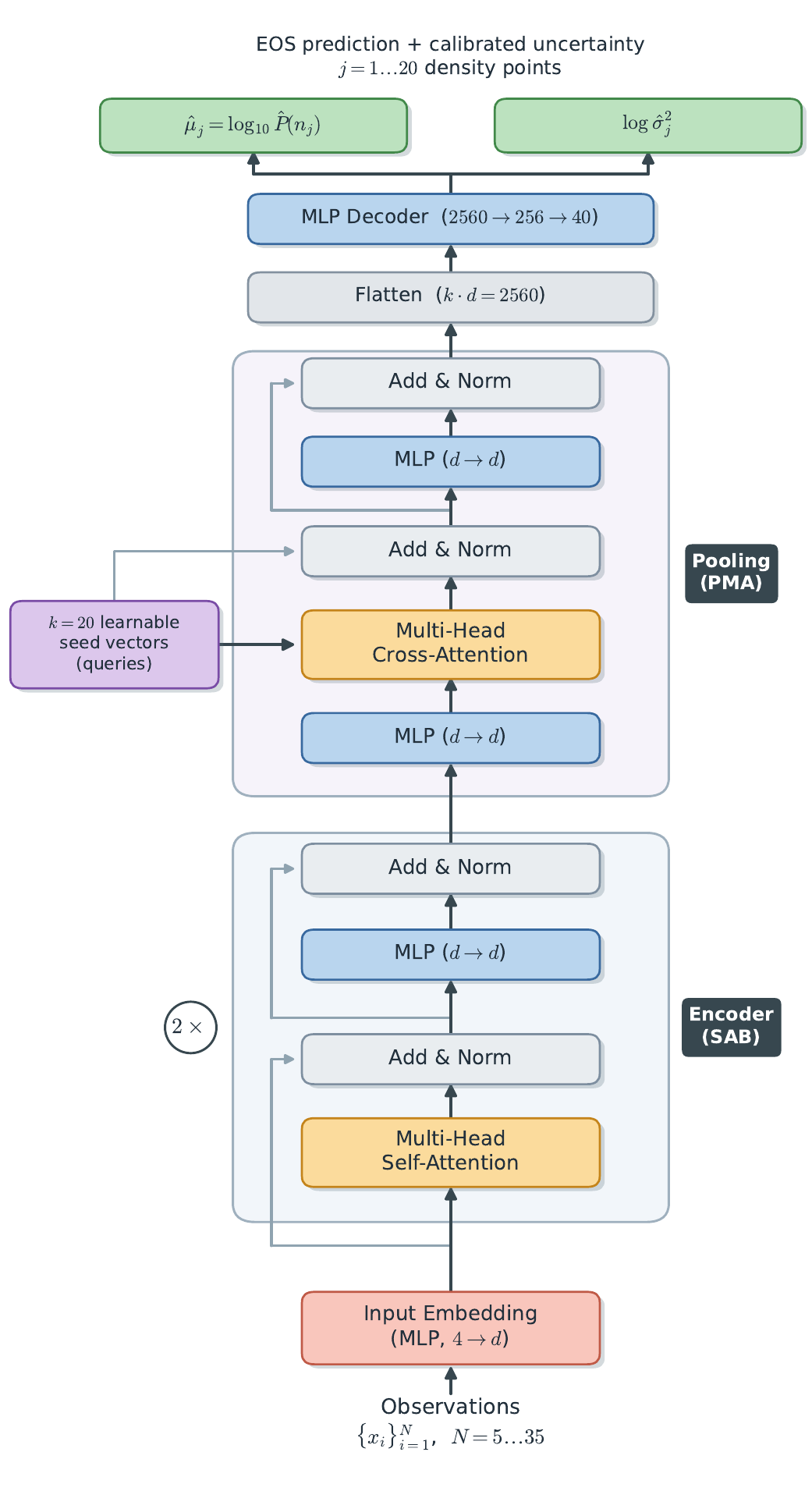}
    \caption{Set Transformer architecture for EoS inference. Observations are
    embedded into $d=128$ dimensions and encoded by two Set Attention Blocks;
    PMA pools the set onto $k=20$ learnable seeds, which are flattened and decoded
    into a mean head ($\log_{10}P$ or $c_s^2$) and a log-variance head at the 20
    grid densities $n_j$. The model is permutation-invariant (no positional
    encoding). See text for details. }
    \label{fig:model}
\end{figure}

The architecture shown in Fig.~\ref{fig:model} is based on the work of Lee et
al.~\cite{lee2019set}. A two-layer {multilayer perceptron (MLP)} embeds
each set element into $d=128$ dimensions. The encoder is composed of two Set
Attention Blocks (SAB), each applying multi-head self-attention over the set
(four heads) followed by {a residual connection and layer normalization,
then an MLP applied to each set element, again with a residual connection and
layer normalization. That MLP is width preserving, mapping $d\to d$.}
Pooling {is performed by} Pooling by Multi-head Attention (PMA) with
$k=20$ learnable seed vectors. The encoded set is first passed through an
MLP of the same kind, which forms the keys and values, while the seeds act as
queries attending to it, producing $k$ pooled vectors. We set the number of
seeds equal to the number of output density points, $k=20$.
{The seeds are not tied to individual densities:} the output of the PMA is
flattened to a $k\cdot d = 2560$ vector and passed to a two-layer {MLP}
decoder that emits, for each of the $20$ densities, a mean $\hat{\mu}_j$ and a
log-variance $\log\hat{\sigma}_j^2$, so each density prediction is a function
of the whole pooled set. This \emph{heteroscedastic} head predicts the mean and
associated uncertainty at each density point. For the pressure target, $\hat{\mu}_j$
predicts $\log_{10}\hat{P}(n_j)$, while for the sound speed target, it predicts
$\hat{c}_s^2(n_j)$ directly. The model has about $10^6$ parameters.

Nothing in the model architecture prescribes which observation informs which output density. Through the attention weights, the network can model nonlinear interactions among all the observations, while the $20$ PMA seeds produce distinct pooled representations of the full encoded set. These representations are jointly decoded into the $20$ density points, so the mapping from stars to densities is learned from the data rather than built in. Moreover, since there is no positional encoding, the same trained network accepts sets of varying size and in any order.

\subsection{Observations, training, and model variants}
\label{sec:train}
During training, the models never see an entire $M(R)$ curve but only mock
observation sets. For a given EoS, we draw a random number of stars
$N\sim\mathcal{U}\{5,\dots,35\}$ once per mini-batch and sample the $N$ masses
uniformly over the EoS's physical branch,
$M_i\sim\mathcal{U}\!\left[0.8\,M_\odot,\,M_{\max}\right]$, where $M_{\max}$ is the
maximum stable mass of that EoS. All figures restrict the analysis to
$M\ge 1\,M_\odot$ while the training range extends below it, so the
$1\,M_\odot$ boundary lies interior to the sampled mass distribution.
The radius is calculated from a cubic spline $R(M)$.
Observational uncertainty is simulated by adding {Gaussian} noise
with standard deviation $\sigma_R$ to each radius. When tidal deformability is
used, a second independent sample set $\Lambda(M)$ is drawn in the same way,
with a relative Gaussian error $\sigma_\Lambda$ (set as a fraction of each
star's $\Lambda$){, resampled until all values are non-negative}.
Masses and noise are resampled every epoch, so each EoS is presented through
many different noisy realizations. Inputs are standardized per channel to zero
mean and unit variance, $z = (x-\mu)/\sigma$, with the mean and standard
deviation computed from the training-set observables. The same statistics are
applied at training and at evaluation. For the models that read tidal
deformability, the $\Lambda$ channel is first mapped to $\log_{10}\Lambda$ and
the statistics are computed on that variable, so the z-score acts on a
well-scaled input.\\

Training minimizes the Gaussian negative log-likelihood (NLL), up to the additive
constant $\tfrac{1}{2}\log 2\pi$,
\begin{equation}
L=\frac{1}{B}\frac{1}{J}\sum_{i=1}^{B}\sum_{j=1}^{J}\tfrac{1}{2}
\left[\log\hat{\sigma}_{ij}^{2}
+\frac{(y_{ij}-\hat{\mu}_{ij})^{2}}{\hat{\sigma}_{ij}^{2}}\right],
\end{equation}
where $i$ runs over the $B$ observation sets in the batch and $j$ over the $J=20$
output density points $n_j$, $\hat{\mu}_{ij}$ and $\log\hat{\sigma}_{ij}^{2}$ are the
two outputs of the network, and $y_{ij}$ is the corresponding true (target) value at
$n_j$.
We monitor the loss value on a {held-out} set every $10$ epochs, and the
checkpoint with the lowest validation loss is kept. A short mean-squared error
(MSE) warm-up of ten epochs was applied to fit the mean before the variance term
{was} enabled. We used the {AdamW} optimizer~\cite{kingma2014adam}
with a learning rate $10^{-3}$ {annealed with a cosine schedule}, weight
decay $10^{-4}$, batch size $64$ {observation sets}, {gradient-norm
clipping to $1.0$}, and trained for {$200$} epochs.
All models are
implemented in PyTorch~\cite{NEURIPS2019_9015}.

We trained a total of twenty models corresponding to ten model configurations, all sharing the same architecture and training procedure. Each configuration was trained independently for the two prediction targets, \(\log_{10}P(n)\) and \(c_s^2(n)\), and the ten configurations differ only in the three design choices listed in Table~\ref{tab:models}.
The input
observations are $R(M)$, $R(M)$ together with $\Lambda(M)$, or $\Lambda(M)$
alone. The observational noise follows one of three scenarios:
$(\sigma_R,\sigma_\Lambda)=(0,0)$, $(0.2 \,\mathrm{km},20\%)$, or
$(0.4\,\mathrm{km},40\%)$. Finally, the model head (see Fig.~\ref{fig:model})
is either the heteroscedastic Gaussian head, which predicts
$\{(\mu_i,\sigma_i)\}$, or a deterministic head with no uncertainty
quantification, {which outputs only $\{\mu_i\}$}. The only trained
model with no uncertainty head is the \texttt{R-MSE}, which was trained
with the mean-squared-error loss. Even when no Gaussian noise is
added to the radius observations, reconstructing the continuous $M(R)$
relation from a finite number of {stars} introduces intrinsic
uncertainty in the inferred EoS. This intrinsic uncertainty is captured by the
\texttt{R-0} model. In summary, both \texttt{R-MSE} and \texttt{R-0} models
use $R(M)$ only as input and their difference is that the \texttt{R-0} model
includes an uncertainty head, predicting $\sigma_j^2$, whereas the
\texttt{R-MSE} model does not. Comparing these two models, therefore, isolates
{the effect} of including the uncertainty head. 
Throughout, $\sigma_R$ and
$\sigma_\Lambda$ denote the noise injected into the observations, while 
$\hat{\sigma}$ denotes the uncertainty predicted by the model. The last
{two columns} of Table~\ref{tab:models} report the mean relative
pressure error and the mean absolute sound speed squared error for each model in the
test set. These results are discussed in Sec.~\ref{sec:results}.
Each configuration was trained once, so the differences between
configurations quoted below carry no estimate of run-to-run variance.
\\

\begin{table*}[!ht]
\caption{
The ten model configurations share the same architecture and are trained on identical data splits. They differ only in (i) the observables provided for each star, (ii) the level of observational noise applied during both training and evaluation, and (iii) the choice of output head. The same ten model configurations are trained for both prediction targets. The noise settings are specified by the pairs $(\sigma_R,\sigma_\Lambda)$, where $\sigma_R$ is the absolute radius uncertainty (km) and $\sigma_\Lambda$ is the fractional uncertainty in each star's tidal deformability, $\Lambda$. The last two columns report the models' performance on the test set via the mean relative pressure error and the mean absolute error of $\cs$ . These metrics were determined using the residuals up to each EoS's $n_{\text{max}}${;} see Fig.~\ref{fig:A0} for details.}
\label{tab:models}
\centering
\renewcommand{\arraystretch}{1.2}
\setlength{\tabcolsep}{6pt}
\begin{tabular}{l l c c l c c}
\toprule
Model & Input data & $\sigma_R$ [km] & $\sigma_\Lambda$ (rel.) & Loss function &
Rel.\ err.\ $P$ [\%] & MAE $c_s^2$ [$c^2$] \\
\midrule
\texttt{R-MSE}     & $R(M)$                 & 0   & ---  & MSE $(\mu)$          & 5.2 & 0.050 \\
\texttt{R-0}       & $R(M)$                 & 0   & ---  & NLL $(\mu,\sigma)$   & 5.1 & 0.051 \\
\texttt{R-$\sigma$}    & $R(M)$            & 0.2 & ---  & NLL $(\mu,\sigma)$   & 6.7 & 0.070 \\
\texttt{R-2$\sigma$}   & $R(M)$            & 0.4 & ---  & NLL $(\mu,\sigma)$   & 8.1 & 0.076 \\
\midrule
\texttt{R$\Lambda$-0}     & $R(M)+\Lambda(M)$ & 0   & 0    & NLL $(\mu,\sigma)$   & 3.3 & 0.044 \\
\texttt{R$\Lambda$-$\sigma$}  & $R(M)+\Lambda(M)$ & 0.2 & 20\% & NLL $(\mu,\sigma)$ & 5.4 & 0.059 \\
\texttt{R$\Lambda$-2$\sigma$} & $R(M)+\Lambda(M)$ & 0.4 & 40\% & NLL $(\mu,\sigma)$ & 6.7 & 0.068 \\
\midrule
\texttt{$\Lambda$-0}     & $\Lambda(M)$       & --- & 0    & NLL $(\mu,\sigma)$   & 4.6 & 0.052 \\
\texttt{$\Lambda$-$\sigma$}  & $\Lambda(M)$    & --- & 20\% & NLL $(\mu,\sigma)$   & 7.5 & 0.072 \\
\texttt{$\Lambda$-2$\sigma$} & $\Lambda(M)$    & --- & 40\% & NLL $(\mu,\sigma)$   & 9.0 & 0.081 \\
\bottomrule
\end{tabular}
\end{table*}

Throughout this work, pressure errors are reported as relative errors (percent) computed in linear pressure space. 
Specifically, since the model predicts \(\log_{10}P\), an error \(d\) in this space corresponds to a multiplicative factor \(10^d\) in pressure. We therefore express the error as a relative percentage, \((10^d-1)\times100\).
  This provides a faithful measure of the error in physical pressure, whereas an absolute error in $\mathrm{MeV}\,\fmc$ would be dominated by the three orders of magnitude variation of $P(n)$ with density rather than the quality of the fit. Sound speed squared errors are reported as mean absolute errors in units of the square of speed of light in vacuum, $c^2$.

\subsection{Data sets}
\label{sec:data}

The two target quantities, the pressure $P$ and the sound speed squared $\cs$ as functions of baryon density $n$, come from two independent EoS ensembles
that share the same $20$-density grid and the same tabulated format. For each
ensemble, the models are trained on one part and evaluated on a separate,
independent test part.

\subsubsection{Pressure (piecewise polytropes ensemble)}
\label{sec:data-poly}

{The pressure target uses an ensemble of EoS parametrized by
piecewise polytropes}~\cite{Ferreira:2024rnf}. Below half the nuclear
saturation density, the crust is fixed to the SLy4 EoS~\cite{sly4}. At higher
densities, the EoS is described by a sequence of connected polytropic segments,
whose adiabatic indices and transition densities are sampled from broad uniform
priors. Only EoS that support a NS with a maximum mass of at least
$2\,\Msun$ and satisfy the {subluminal} sound speed constraint are
retained, {leaving} $40\,435$ {EoS, split into}
$36\,405$ {training and} $4\,030$ {test EoS.} The
regression target is $\log_{10}P(n)$, as the pressure spans
about three orders of magnitude across the density grid,
which makes the regression better conditioned in logarithmic
space. Since the crust is common to the whole ensemble and the density grid
begins at $0.10\,\fmc$, part of the accuracy at the lowest grid points is
inherited from this fixed low-density input rather than learned from the
observations.
\subsubsection{Sound speed (GP ensemble)}
\label{sec:data-gp}

The sound speed target is trained on an independent nonparametric ensemble in
which the sound speed squared $\cs(n)$ is drawn from a GP ~\cite{Annala2023,Gorda2023,komoltsev_2023_10101447} and reweighted by astrophysical
constraints {(X-ray radius measurements, the} $2\,\Msun$
{radio-timing constraints, and perturbative-QCD behavior at high
density)}. The full prior ensemble contains $120\,000$ EoS; restricting to
those that reach at least $2\,M_\odot$
{($M_{\rm max} \geq 2\,M_\odot$) leaves $77\,983$ of them, split}
$70\,185/7\,798$ into training and test sets. Here, the target is the raw
$\cs(n)$, which is bounded in $(0,1)$ and regressed directly,
{without a logarithmic transformation.}

%======================================================================
\section{Results}
\label{sec:results}
%======================================================================

\begin{figure*}[!t]
    \centering
    \includegraphics[width=0.95\linewidth]{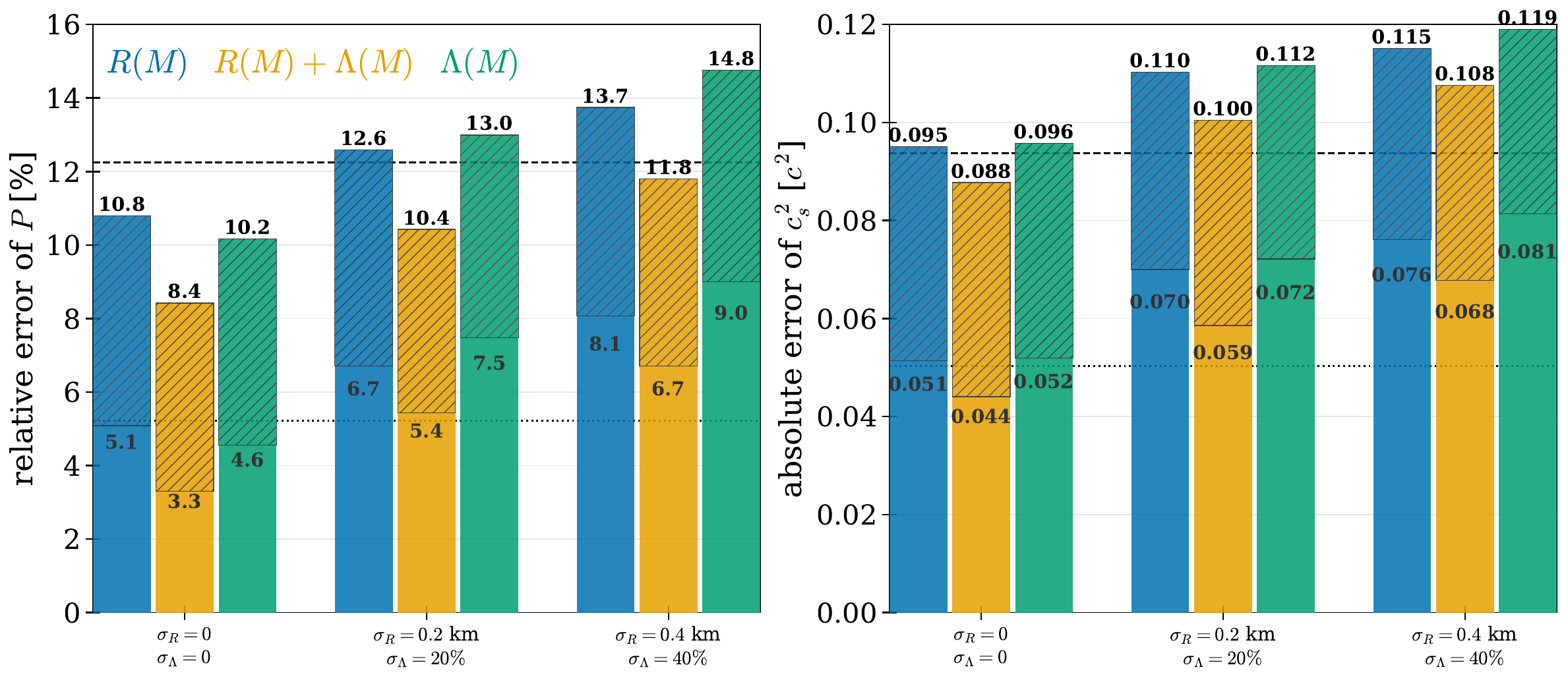}
    \caption{Accuracy of the ten model configurations evaluated on the test
    set, with the baseline \texttt{R-MSE} model shown as horizontal lines.
    \emph{Left}: mean relative pressure error
    $\left[|\hat P-P|/P\right]$ $[\%]$ 
    over the
    $4\,030$ polytropic test EoS, evaluated with one observation
    set per EoS, each containing 5--35 randomly drawn stars. \emph{Right}:
    mean absolute error of
    $c_s^2$ for the $7\,798$ GP test EoS. The colored bars
    denote the input observables: $R(M)$ (blue), $R(M)+\Lambda(M)$ (orange), and
    $\Lambda(M)$ (green); bar groups correspond to the observational-noise
    scenarios, with the corresponding $\sigma_R$ and $\sigma_\Lambda$ values
    given on the abscissa. The total bar height is the error over the full
    density range $n\le 1.10~\mathrm{fm}^{-3}$, the solid part is the error
    restricted to $n\le n_{\text{max}}$ for each EoS, and the hatched part is
    the additional error from the extrapolation region $n>n_{\text{max}}$. The
    dashed line marks the baseline $R(M)$ model trained with an MSE loss and no
    uncertainty head, evaluated over the full grid, and the dotted line marks
    the same baseline restricted to $n\le n_{\text{max}}$.}
    \label{fig:A0}
\end{figure*}

% Figure~\ref{fig:A0} summarizes the performance of all twenty trained models on the held-out test EoS: three input modes crossed with
% three noise levels, plus a noise-free baseline for each of the two
% targets.
Figure~\ref{fig:A0} summarizes the predictive accuracy of the ten model configurations on the test EoSs. The left panel shows the mean relative error in the reconstructed pressure, while the right panel shows the mean absolute error for \(c_s^2\). The three input modes are evaluated under the three observational-noise scenarios, and the noise-free \texttt{R-MSE} is used as a baseline and shown separately as horizontal reference lines.

For each model, the errors are evaluated in two ways: over the full density grid
$n\le 1.10~\mathrm{fm}^{-3}$ {(total bar height)} or by using only
the density points up to $n_{\text{max}}$ of each EoS {(solid
bars)}, where $n_{\text{max}}$ is the central baryon density of the
maximum-mass configuration of that EoS, i.e., the highest density reached by any stable star. The former unavoidably includes the extrapolation region
$n > n_{\text{max}}$, which no stable star can probe{;} the latter
measures the error over the physically reachable range
$n \le n_{\text{max}}$. The hatched portion of each bar therefore quantifies
the extra error incurred purely by extrapolation. In the discussion below, we
quote the solid-bar values, i.e.\ the errors restricted to the range
$n \le n_{\text{max}}$ {accessible to stable stars}.
Three conclusions follow.
First, augmenting the mass--radius input data with tidal deformability
improves pressure reconstruction in all noise scenarios. For example, at
$\sigma_R=0.2$~km, the mean relative pressure error up to $n_{\text{max}}$
decreases from $6.7\%$ to $5.4\%$ (from $12.6\%$ to
{$10.4\%$} over the full grid), i.e.\ the gain
{persists over the full grid as well}.
Second, the presence of an uncertainty head {(see}
Fig.~\ref{fig:model}{)} in the model architecture does not reduce
predictive accuracy. The \texttt{R-0}
model (first blue bar) achieves a mean relative pressure error of $5.1\%$ up
to $n_{\text{max}}$, compared with $5.2\%$ for the \texttt{R-MSE} model
(dotted line), showing that an uncertainty estimation is effectively
obtained at no cost in {mean accuracy}. The models that
incorporate tidal deformability perform even better, with \texttt{R$\Lambda$-0}
reaching $3.3\%$ and \texttt{$\Lambda$-0} $4.6\%$. These gains arise from the
increased information content of the input vector rather than from the
heteroscedastic output head itself.
Third, model performance degrades smoothly as observational noise increases.
In the noiseless limit, {the mean absolute error of} $c_s^2$ up to
$n_{\text{max}}$ {is} $0.044$--$0.052$ {across input
modes, with the combined $R+\Lambda$ model already the most accurate}
($0.044$). Differences between the input configurations become more pronounced
once observational noise is introduced. At $\sigma_R = 0.4$~km the absolute
error rises to $0.068$ for $R+\Lambda$ versus $0.076$--$0.081$ for the
single-channel inputs, and the combined $R+\Lambda$ input consistently
provides the best performance.

\subsection{Pressure}
\label{sec:pressure}

\subsubsection{Reconstruction}
\begin{figure*}[!ht]
    \centering
    \includegraphics[width=0.9\linewidth]{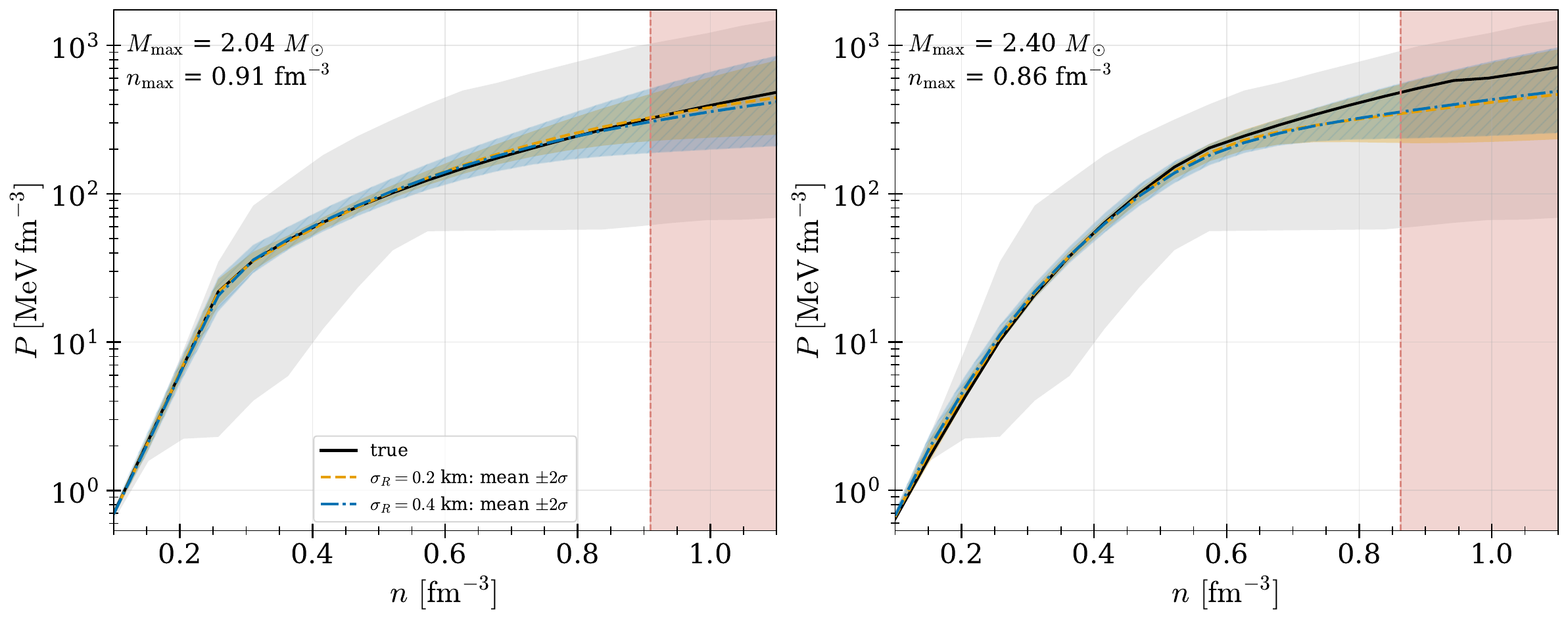}
\caption{Pressure reconstruction from a single simulated observation set for two polytropic EoS. The true pressure is shown as a solid black line. The
predicted means and $\pm2\sigma$ bands {are overlaid for the
models trained and evaluated at} radius noise of $\sigma_R=0.2$~km (orange,
dashed mean, solid band,  $R-\sigma$ model) and
$\sigma_R=0.4$~km (blue, dash-dotted mean, hatched band, $R-2\sigma$ model). Each
inference uses a set of $N=15$ observations. The vertical
axis is logarithmic. The gray region indicates the pressure range covered by
the training dataset, while the red-shaded region indicates densities above the
maximum density supported by the respective EoS, $n>n_{\mathrm{max}}$.}
    \label{fig:A2}
\end{figure*}

Figure~\ref{fig:A2} shows the reconstructions of two test EoSs, a soft EoS with $M_{\mathrm{max}} = 2.04\,M_{\odot}$ (left) and a stiff EoS with $M_{\mathrm{max}} = 2.40\,M_{\odot}$ (right), from
single simulated observation sets.
Each model was applied to its own observation set of $N=15$
simulated stars drawn from the EoS $M(R)$ solution, with the corresponding radius
noise ($R-\sigma$ model with $\sigma_R=0.2$~km and $R-2\sigma$ model with $\sigma_R=0.4$~km).
The true pressure lies within the $\pm 2\hat{\sigma}$ bands of both models
throughout the supported density range,
$n \leq n_{\mathrm{max}}$, and the two mean predictions nearly coincide.
{The wider band of the \texttt{R-$2\sigma$} model reflects its
doubled input noise: the predictive uncertainty broadens substantially, while
the mean reconstruction changes only modestly.}

\subsubsection{Robustness to observation count} 

\begin{figure}[!ht]
    \centering
    \includegraphics[width=0.5\linewidth]{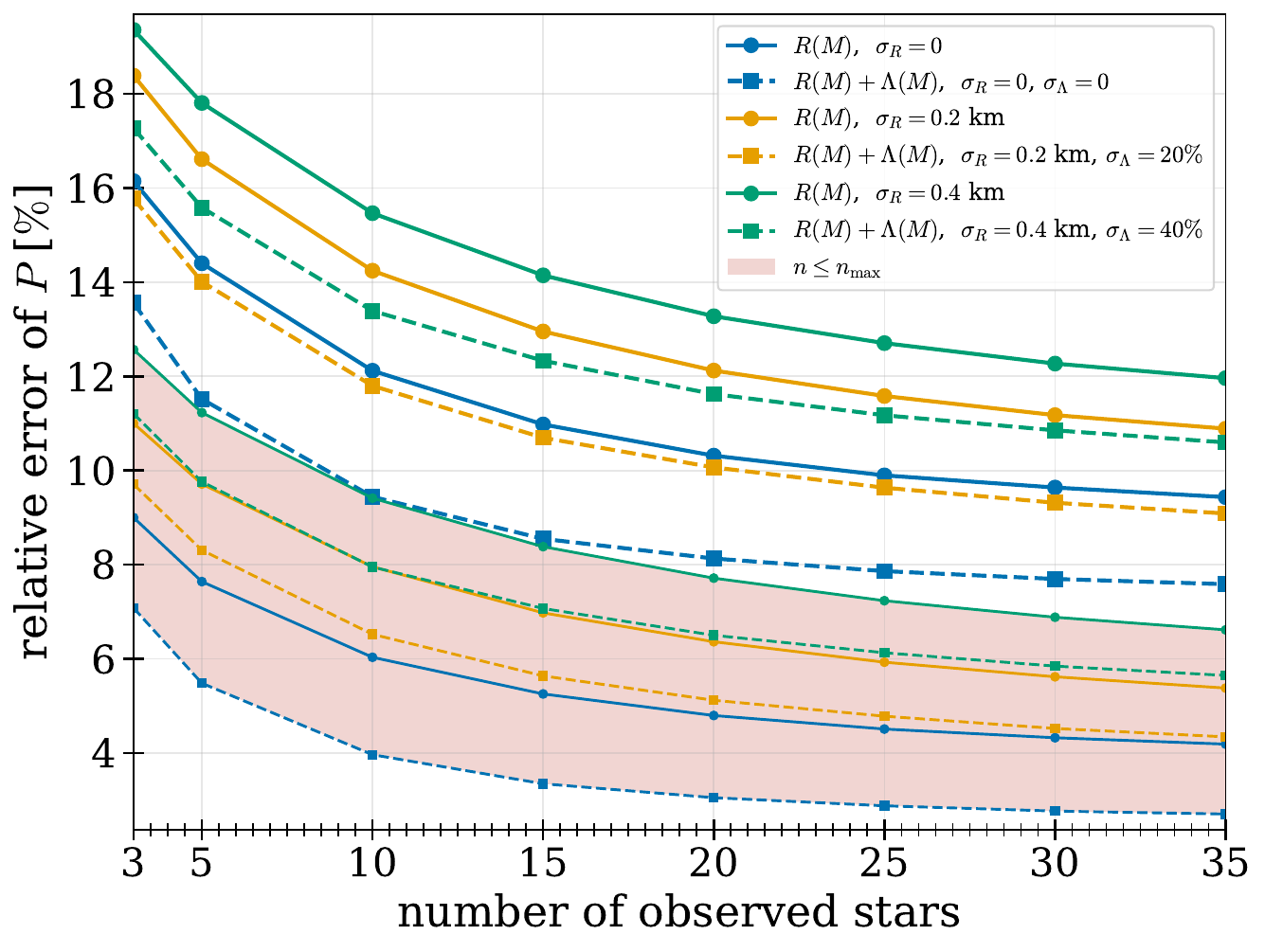}
\caption{Mean relative pressure error, $\langle |\hat P-P|/P\rangle\times100$, on the test set as a function
of the number of observations. Colors denote the noise scenario:
{$\sigma_R=0$ (blue), $\sigma_R=0.2\,\mathrm{km}$ (orange), and
$\sigma_R=0.4\,\mathrm{km}$ (green).} 
Solid lines with circles correspond to
$R(M)$, while dashed lines with squares correspond to $R(M)+\Lambda(M)$. Thick curves use the complete
density grid, including the extrapolation region; thin curves of the same line style and the
red-shaded band are restricted to the physically supported region,
$n\leq n_{\mathrm{max}}$. All curves are averaged over the $4\,030$ test EoS and 20 independent
    observation sets for every
test EoS.}
    \label{fig:A3}
\end{figure}

A central question is how many observations are needed to achieve a given
accuracy in the EoS reconstruction. Figure~\ref{fig:A3} shows the mean relative pressure error as a function of the number of observed
stars $N$, from 3 to 35, with $N=3$ lying below the training range
$N=5$--$35$, for three noise scenarios and two input modes,
$R(M)$ (solid  lines with circles) and $R(M)+\Lambda(M)$ (dashed lines with squares). For the combined $R(M)+\Lambda(M)$ model, $N$ counts the
observations of each type, so the set contains $2N$ elements in total.
The thick curves cover the
complete density grid, including the extrapolation region above the maximum
density supported by each EoS.
The thin curves inside the red-shaded band
show the corresponding errors restricted to the physically supported region,
$n\leq n_{\mathrm{max}}$. 

For each value of $N$, 20 independently generated observation sets were used
for each of the {$4\,030$} test EoS, and the error was averaged
over both the EoS ensemble and the observation set realizations. The
$R(M)+\Lambda(M)$ mode generally yields a lower error than $R(M)$ across the
noise scenarios and observation counts, {although} in the noisy
cases, the $\Lambda$ channel carries its own relative observational noise
($\sigma_\Lambda=20\%$ and $40\%$).
The supported-density curves show the same overall trend as the full-grid
curves, while excluding the extrapolation region
{consistently} reduces the measured reconstruction error. This
indicates that part of the error in the full-grid evaluation arises from
predictions beyond the density range physically supported by the individual
EoS. The additional information provided by $\Lambda(M)$ can compensate for
its observational noise: for $\sigma_R=0.2\,\mathrm{km}$, ten stars with the
$\Lambda$ channel achieve a mean relative error of {$11.8\%$},
compared with {$12.1\%$} for twenty stars without it --- the same total
number of set elements.
Restricting the error to $n \leq n_{\mathrm{max}}$,
the compensation persists: ten stars with the $\Lambda$ channel reach
$6.5\%$, matching the $6.4\%$ of twenty stars without it, and doubling
the observational noise on both channels ($R\Lambda$, $\sigma_R = 0.4\,
\mathrm{km}$, $\sigma_\Lambda = 40\%$) leaves the accuracy essentially
unchanged relative to the $R(M)$ model at $\sigma_R = 0.2\,\mathrm{km}$
($7.9\%$ vs.\ $8.1\%$ at ten stars, $6.5\%$ vs.\ $6.4\%$ at twenty).
This holds for all six configurations (three noise scenarios
$\times$ two input modes): the error decreases over the explored range of
observation-set sizes $N$, although the improvement becomes smaller
as $N$ increases. No saturation is observed within the range shown.
Although $N=3$ lies below the training range, the model still produces
finite and physically meaningful reconstructions, with errors that remain
moderate and decrease as the number of observations increases. This
provides a direct test of the model's generalization to set sizes outside
the training range.

\subsubsection{Density sensitivity to \texorpdfstring{$R(M)$}{R(M)} and
\texorpdfstring{$\Lambda(M)$}{Lambda(M)} observations}

For a fixed EoS and observation set, we quantify how strongly the predicted
pressure at density $n$ depends on each observed star, separately through
each input observable. For star $j$ and density $n_i$, we define the
per-channel sensitivities
\begin{equation}
s_j^M(n_i) = \left|\frac{\partial \hat{P}(n_i)}{\partial M_j}\right|,
\qquad
s_j^R(n_i) = \left|\frac{\partial \hat{P}(n_i)}{\partial R_j}\right|,
\label{eq:sensitivity}
\end{equation}
computed as the absolute Jacobian of the model output with respect to the physical observables. 

We normalize the sensitivities across all stars in the set at each density point:
\begin{equation}
\widetilde{s}_j^M(n_i)
=
\frac{s_j^M(n_i)}
{\sum_{j'} s_{j'}^M(n_i)},
\qquad
\widetilde{s}_j^R(n_i)
=
\frac{s_j^R(n_i)}
{\sum_{j'} s_{j'}^R(n_i)}.
\label{eq:sensitivity_norm}
\end{equation}
Thus,
\begin{equation}
\sum_j \widetilde{s}_j^M(n_i)
=
\sum_j \widetilde{s}_j^R(n_i)
=
1.
\end{equation}

\begin{figure*}[!t]
    \centering
    \includegraphics[width=0.8\linewidth]{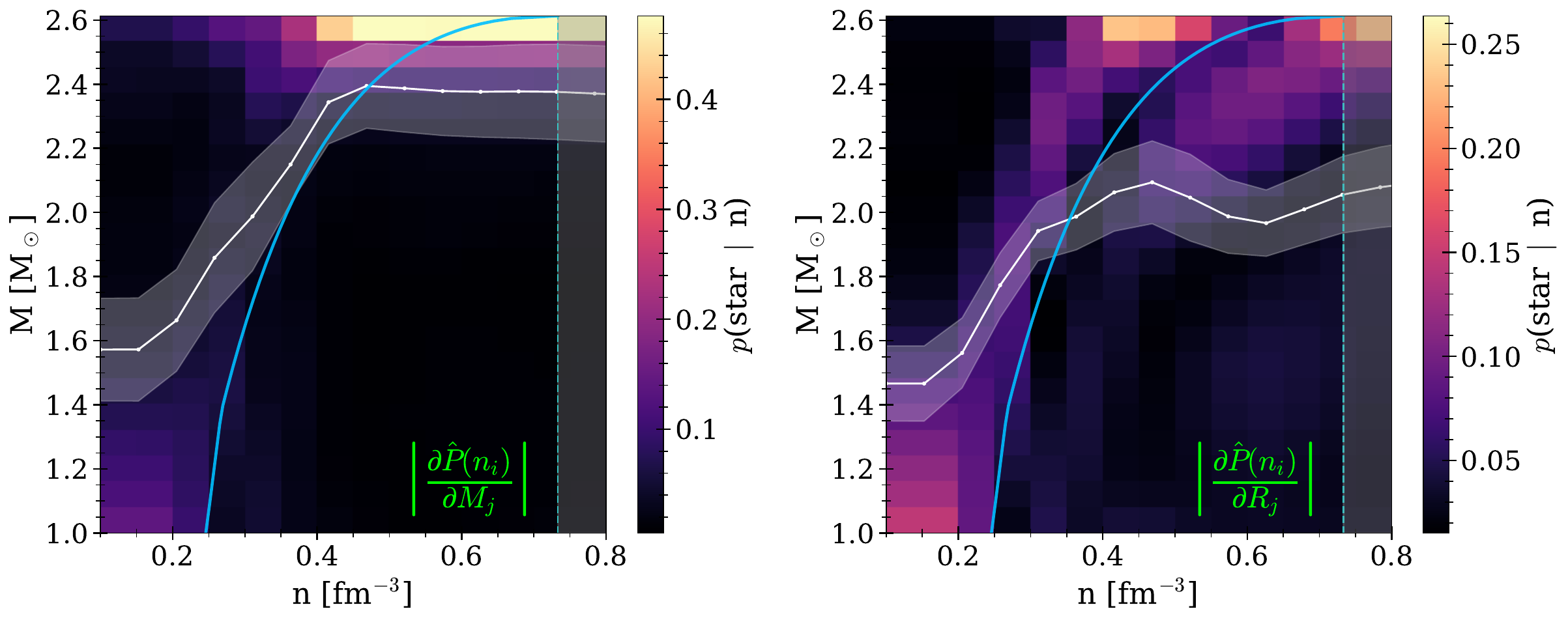}
\caption{
Sensitivity maps of the \texttt{R-$\sigma$} model ($R(M)$ input data,
$\sigma_R = 0.2\,\mathrm{km}$) to the mass and radius of the observed stars
for the pressure target, evaluated for a reference polytropic EoS with
$M_{\mathrm{max}} = 2.61\,M_{\odot}$ and $n_{\mathrm{max}} = 0.73\,
\mathrm{fm}^{-3}$. Each panel shows one input channel in its own physical
units: the mass channel $|\partial\hat{P}(n_i)/\partial M_j|$ (left) and
the radius channel $|\partial\hat{P}(n_i)/\partial R_j|$ (right), with the
derivatives taken with respect to the raw physical observables. The colour
encodes $p(\text{star}\mid n)$, the share of that channel's sensitivity
carried by stars of each mass at density $n$. The sensitivity-weighted
mean mass, i.e.\ the centre of mass of $p(\text{star}\mid n)$, and its
$1\sigma$ spread are shown in white, and the EoS's central-density--mass
relation $n_c(M)$ in cyan. The dashed line and gray band mark the
extrapolation region $n > n_{\mathrm{max}}$. See text for details.}
    \label{fig:A4}
\end{figure*}

Figure~\ref{fig:A4} shows the two resulting sensitivity maps for a fixed
polytropic EoS with $M_{\mathrm{max}}=2.61\,M_{\odot}$ and
$n_{\mathrm{max}}=0.73\,\mathrm{fm}^{-3}$. The left panel shows the mass
channel $s^M$ and the right panel shows the radius channel $s^R$, each in
its own physical units. The maps were obtained from 6000 observation sets of
$N=5$ to 35 stars, with masses sampled uniformly from
$[1,M_{\mathrm{max}}]\,M_\odot$ and radius noise $\sigma_R=0.2\,\mathrm{km}$.
The colour in each panel is the quantity $p(\text{star}\mid n)$, the share
of that channel's sensitivity carried by stars of each mass at a given
density. For every observation set, the stars are sorted into 20 equal mass bins over
$[1,M_{\mathrm{max}}]\,M_\odot$ and the normalized sensitivities of
Eq.~\eqref{eq:sensitivity_norm} are added within each bin. Averaging these bin
shares over the 6000 sets yields $p(\text{star}\mid n_i)$, a probability
distribution over stellar mass at each density. The white line traces the
centre of mass of $p(\text{star}\mid n)$, the sensitivity-weighted mean
mass, together with its $1\sigma$ spread. The cyan curve shows the
central-density--mass relation $n_c(M)$ for the analyzed EoS. The dashed
line and the gray shading mark the extrapolation region
$n>n_{\mathrm{max}}$. Because each column is normalized, $p(\text{star}\mid
n)$ identifies which stars shape $\hat{P}(n)$, but not how strongly that
density is constrained. It reflects the influence of stars of each mass,
not their abundance in the sets, which is uniform by construction.

In both panels, the sensitivity migrates from lighter to heavier stars as
the reconstructed density increases, following $n_c(M)$ at low to
intermediate densities. The two channels then show distinct behaviors. In the mass
channel, the informative mass climbs steadily and saturates among the
heaviest stars of the sequence, just below $M_{\mathrm{max}}$. There, the
stable $M(R)$ sequence approaches a horizontal tangent, i.e.\ $dM/dR=0$, and terminates, so no stable star heavier
than $M_{\mathrm{max}}$ exists to probe higher central densities. 
A second, distinct feature appears lower on the sequence. At the
point of maximum radius, $M_{R\text{max}} \simeq 2.04\,M_{\odot}$ for this
EoS, the $M(R)$ curve has a vertical tangent, i.e. $dR/dM = 0$. A change in mass at fixed radius therefore displaces the star almost along its own $M(R)$ curve rather than across the family of curves generated by different EoS: the perturbed observation remains compatible with the same EoS, and the mass channel loses nearly all of its sensitivity there.
In the radius channel, the roles are reversed. The informative mass stays close to $M_{R\text{max}}$ over most of the density range, while the radii of light stars carry a large share of the sensitivity at low density, where a
light star's radius is fixed by the low density part of the EoS. Unlike
the mass channel, the radius keeps contributing at every density,
including beyond $n_{\mathrm{max}}$, because it constrains the EoS over
the whole density range a star spans rather than at a single density. The
two channels are thus complementary. The mass localizes the inference at
the density a star actually reaches, while the radius provides a global
constraint that never fully fades.

\begin{figure*}[!ht]
    \centering
    \includegraphics[width=0.8\linewidth]{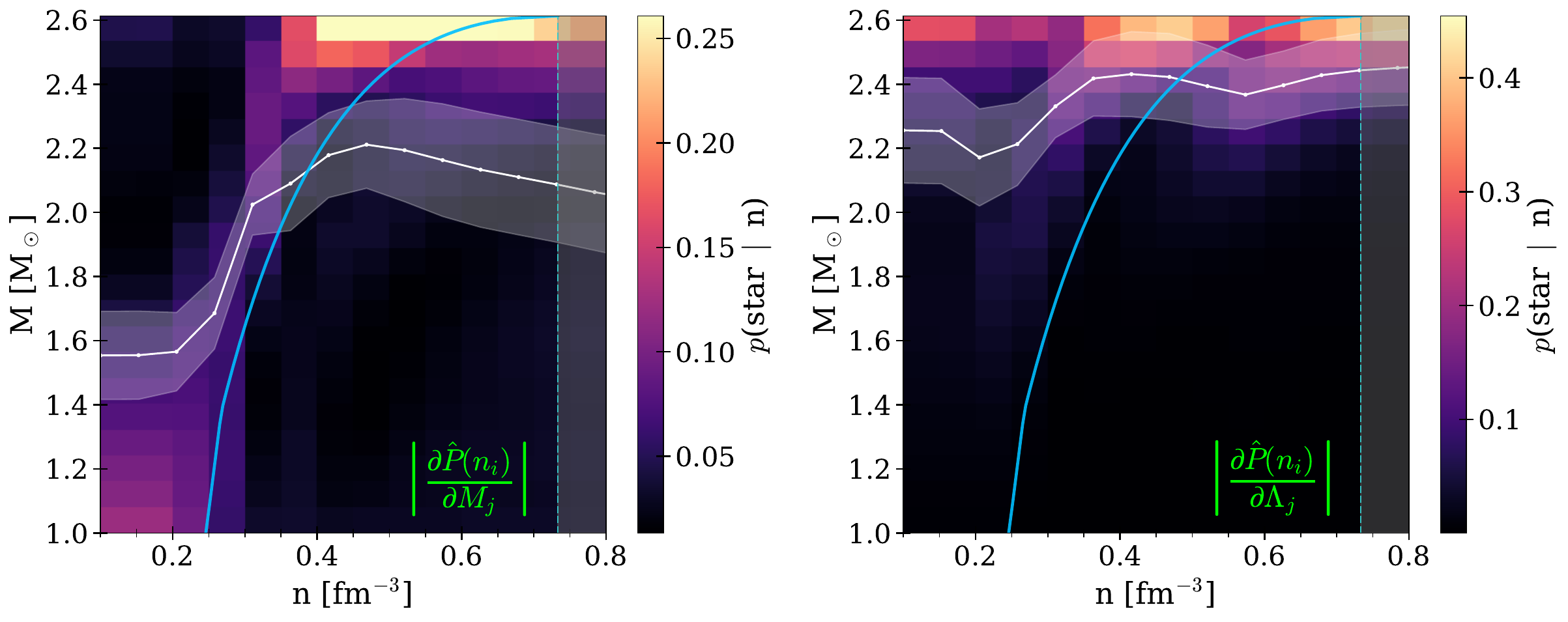}
\caption{Same construction as Fig.~\ref{fig:A4} (same reference EoS, probe
and plotting conventions), now for the \texttt{$\Lambda$-$\sigma$} model,
trained on $\Lambda(M)$ observations with $\sigma_\Lambda = 20\%$ relative
noise instead of the $R(M)$ observations of the \texttt{R-$\sigma$} model.
The right panel shows the tidal-deformability channel
$|\partial\hat{P}(n_i)/\partial\Lambda_j|$ in place of the radius channel,
the derivative being taken with respect to the physical $\Lambda$.}
    \label{fig:A5}
\end{figure*}

The pattern is not specific to the radius observable. Figure~\ref{fig:A5}
shows the corresponding sensitivity maps for the \texttt{$\Lambda$-$\sigma$}
model, trained on $\Lambda(M)$ input data with $\sigma_\Lambda=20\%$
relative noise and evaluated on the same reference EoS as
Fig.~\ref{fig:A4}. The left panel again shows the mass channel
$|\partial\hat{P}/\partial M|$, and the right panel now shows the
tidal-deformability channel $|\partial\hat{P}/\partial\Lambda|$, the
derivative being taken with respect to the physical $\Lambda$ (the model
input is $\log_{10}\Lambda$). 
The mass channel still
traces a diagonal band, but a weaker one: its informative mass rises with
density, then drifts back toward lower masses at the highest densities.

The $\Lambda$
channel is instead concentrated on the most massive stars at every density,
with an informative mass that stays between $\approx 2.4$ and
$\approx 2.5\,M_{\odot}$ at all densities.
This concentration is the tidal analog of the mass channel saturating at
$M_{\mathrm{max}}$, and it follows from expressing the sensitivity per unit
of $\Lambda$. Along the mass--radius curve, $\Lambda$ itself falls by about
three orders of magnitude, from $\approx 4.6\times10^{3}$ at $1\,M_\odot$
to $\approx 5$ at $M_{\mathrm{max}}$. A fixed absolute change of $\Lambda$
is therefore a substantial fractional change for a massive star and a
negligible one for a light star.
The tidal deformability is a global observable. It is an integrated
response of the star, sensitive to the EoS over the entire interior rather
than at a single density, and it decreases monotonically with mass. The
sensitivity to any given star's $\Lambda$ therefore persists at all
densities, including the extrapolation region, so the network relies on it
as a global probe of the EoS, complementary to the density-localized mass
channel.

\subsubsection{Compactness}

\begin{figure*}[!t]
    \centering
    \includegraphics[width=0.9\linewidth]{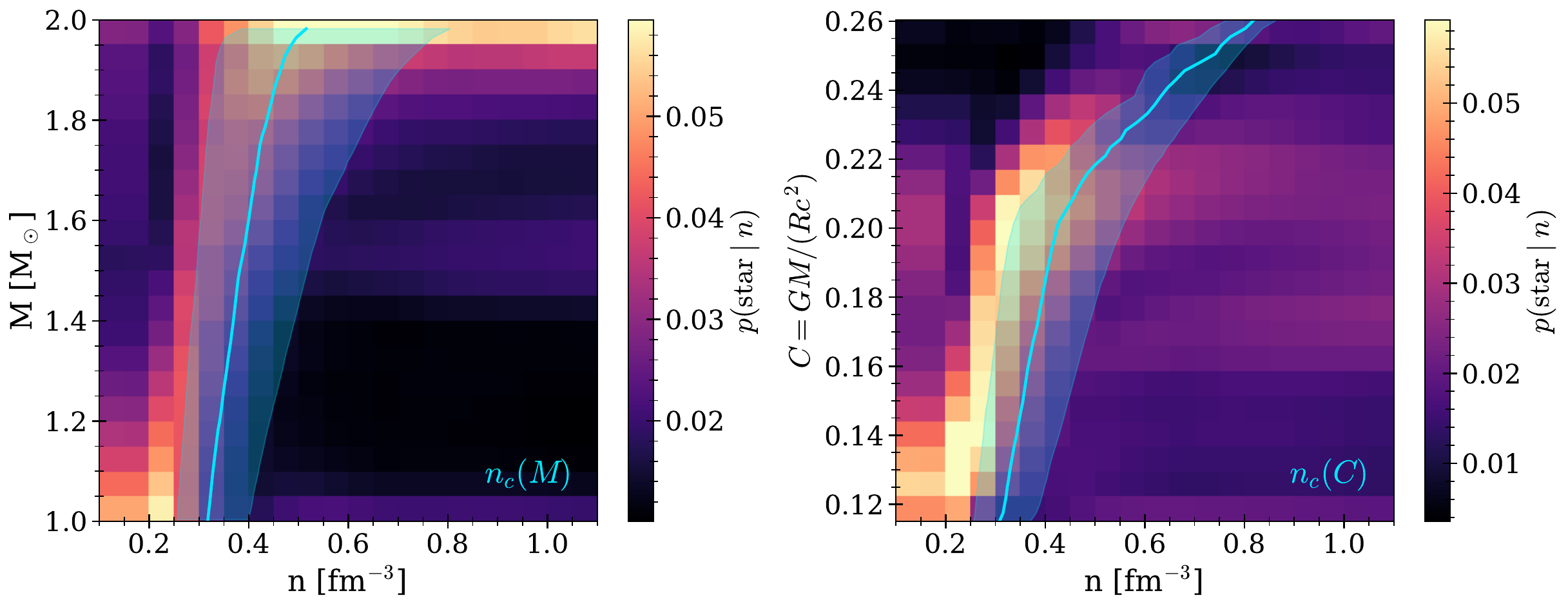}
\caption{Ensemble-averaged sensitivity of the \texttt{R-$\sigma$} model
($R(M)$ input data, $\sigma_R = 0.2\,\mathrm{km}$) for the pressure target,
with the two channels combined through Eq.~\eqref{eq:sensitivityC} and
averaged over the $4\,030$ polytropic test EoS. The colour encodes
$p(\text{star}\mid n)$, the share of the combined sensitivity carried by
stars of a given mass (left) or compactness (right) at density $n$. The
cyan curve and band show the median and 10th--90th-percentile range of the
central-density relations $n_c(M)$ and $n_c(C)$ across the ensemble.}
    \label{fig:A6}
\end{figure*}

The maps of Figs.~\ref{fig:A4} and~\ref{fig:A5} are expressed in
stellar-mass coordinates, although each EoS has its own
central-density--mass relation $n_c(M)$, so stars of the same mass probe
different central densities in different EoS. 
Figure~\ref{fig:A6} tests whether the compactness
$C = GM/(Rc^2)$ provides a more universal coordinate.
The probe combines the two channels of Eq.~\eqref{eq:sensitivity} into a
single, dimensionless sensitivity. For star $j$ and density $n_i$,
\begin{equation}
s_j(n_i) \;=\; M_j\, s_j^M(n_i) + R_j\, s_j^R(n_i)
\;=\; \left|\frac{\partial \hat{P}(n_i)}{\partial \ln M_j}\right|
     + \left|\frac{\partial \hat{P}(n_i)}{\partial \ln R_j}\right|,
\label{eq:sensitivityC}
\end{equation}
the response of the prediction to a \emph{fractional} change of each
observable. 

Each of the $4\,030$ polytropic test EoS is probed with a stratified set of
$N=20$ stars, one per equal mass bin, spanning its own physical mass range. For each EoS, the sensitivities are
normalized over the full set, so each column reports the share of the total sensitivity carried
by stars inside the plotted window. Averaged over the ensemble, the result
is the channel-combined analog of $p(\text{star}\mid n)$: the left panel
of Fig.~\ref{fig:A6} shows it in mass coordinates, and the right panel
re-bins the same sensitivities by compactness. The cyan curve and band show
the median and 10th--90th percentile range of the true relations $n_c(M)$
and $n_c(C)$.

Within the $[1,2]\,M_\odot$ window, the spread of central densities among
stars of the same mass drops from $0.090$ to $0.064\,\mathrm{fm}^{-3}$ when
stars are grouped by compactness instead, about $30\%$ smaller.  Stars of the same compactness thus
have more similar central densities than stars of the same mass. This narrowing of the central density spread, computed directly from the
EoS ensemble, is the decisive quantity of the figure, and
the model map serves only to illustrate it. In mass coordinates, each EoS
has its own central density curve $n_c(M)$, and the band can only run
loosely among them. Re-binned by compactness, the same sensitivities
concentrate around the single universal $n_c(C)$ relation.
The band does not lie exactly on the curve. A star's observables
constrain the EoS at all densities up to its own central density, so its
sensitivity is spread over densities below $n_c$, and the band sits slightly
above the universal relation, by $\simeq 0.02$--$0.03$ in $C$.

\subsection{Sound speed}
\label{sec:soundspeed}

\subsubsection{Reconstruction}
\begin{figure*}[!t]
    \centering
    \includegraphics[width=0.9\linewidth]{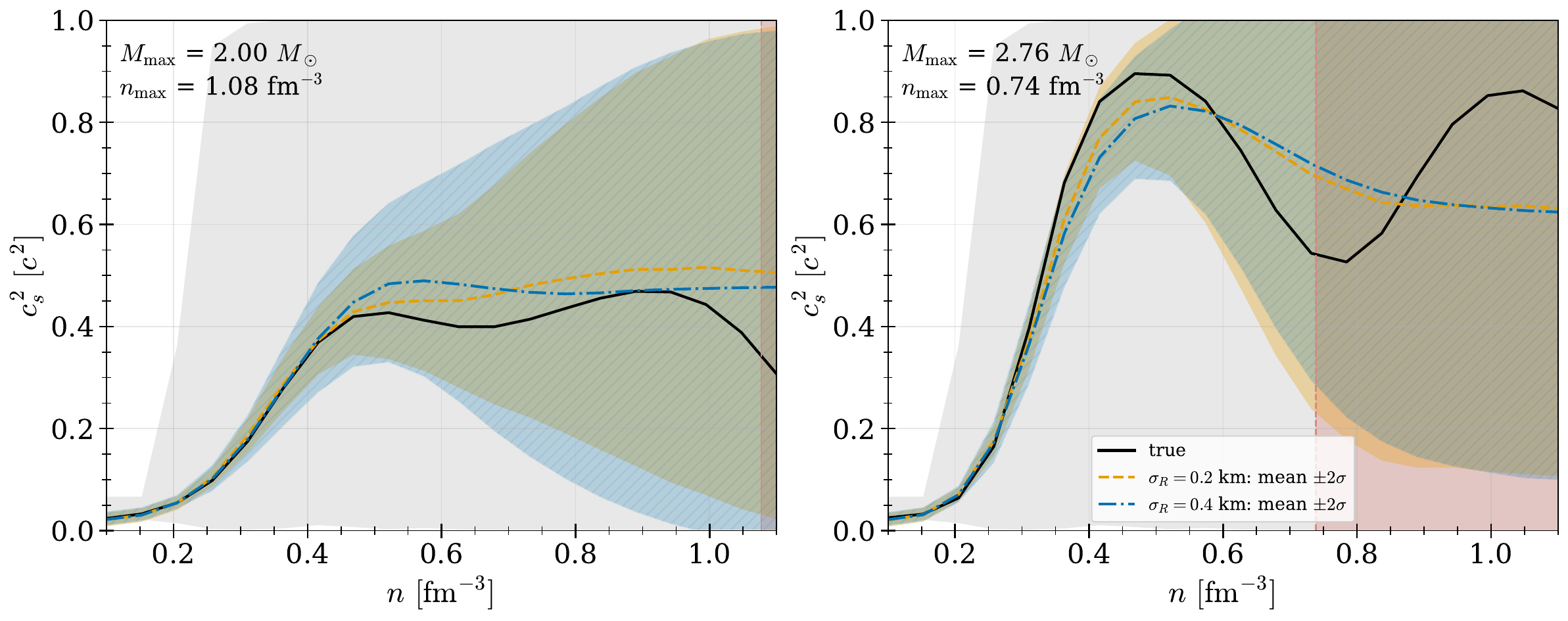}
\caption{Sound speed squared reconstruction from a single simulated observation set
for a soft (left) and a stiff (right) EoS. The true $c_s^2$ is
shown in black. Each
inference uses $N=15$ observations. The predicted means and
$\pm2\sigma$ bands for models trained with and applied to radius noise
$\sigma_R=0.2\,\mathrm{km}$ (orange, dashed mean, solid band) and
$\sigma_R=0.4\,\mathrm{km}$ (blue, dash-dotted mean, hatched band) are
overlaid. The gray region indicates {the $c_s^2$ range covered by
the training dataset}, while the red-shaded
region marks densities above the maximum density supported by the respective
EoS, $n>n_{\mathrm{max}}$.}
    \label{fig:V2}
\end{figure*}

Figure~\ref{fig:V2} shows the sound speed squared reconstructions of two unseen test
EoS from single simulated observation sets. The left panel shows a
soft EoS with $M_{\mathrm{max}}=2.0\,M_{\odot}$, while the right panel shows a stiff
EoS with $M_{\mathrm{max}}=2.76\,M_{\odot}$. The \texttt{R-$\sigma$} model
($\sigma_R=0.2\,\mathrm{km}$, orange dashed mean) and the
\texttt{R-$2\sigma$} model ($\sigma_R=0.4\,\mathrm{km}$, blue dash-dotted
mean) each use one observation set with the corresponding radius noise.
The true, non-monotonic $c_s^2$ profile lies within the $\pm2\hat{\sigma}$
bands of both models throughout the supported density range. For the soft EoS,
the uncertainty grows toward high densities because its light stars mainly
constrain the low-density region. The model therefore
{predicts increasing uncertainty} while retaining a reasonable
mean prediction.  For the stiff EoS, the uncertainty remains
relatively small below $0.5\,\mathrm{fm}^{-3}$ and increases afterward due to
the {non-monotonic} behavior. {The mean predictions
nevertheless track the true profile.} In this example, doubling the
radius noise primarily broadens the uncertainty band, while the two mean
reconstructions remain similar, which is similar to what we obtained in our previous work \cite{Carvalho:2023ele}.

\subsubsection{Robustness to observation count}
\begin{figure}[!ht]
    \centering
    \includegraphics[width=0.5\linewidth]{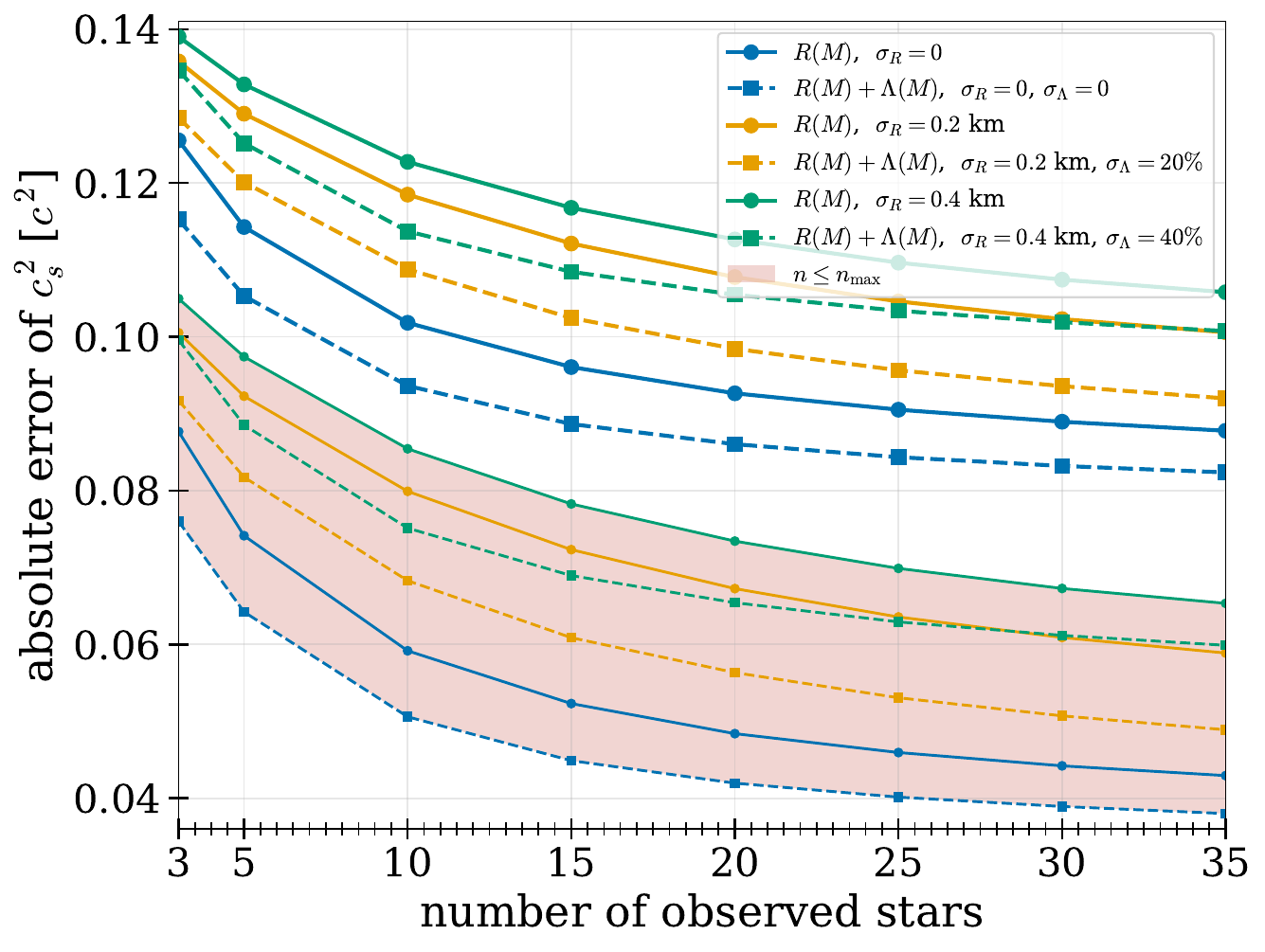}
    \caption{Mean absolute error of the reconstructed sound speed,
    $\langle |\hat{c_{s}}^2-c_{s}^2|\rangle$ in units of $c^2$, on
    the test set as a function of the number of observations. Colors denote the
    noise scenario: {$\sigma_R=0$ (blue), $\sigma_R=0.2\,\mathrm{km}$
    (orange), and $\sigma_R=0.4\,\mathrm{km}$ (green).} Solid lines with circles correspond to
$R(M)$, while dashed lines with squares correspond to $R(M)+\Lambda(M)$. Thick curves use the complete
density grid, including the extrapolation region; thin curves of the same line style and the
red-shaded band are restricted to the physically supported region,
$n\leq n_{\mathrm{max}}$. All curves are averaged over the $7\,798$ test
    EoS and 20 independent observation sets for every
test EoS.}
    \label{fig:V3}
\end{figure}

Figure~\ref{fig:V3} shows the mean absolute error of the reconstructed sound speed, $|c_s^2-\hat{c}_s^2|$, where $\hat{c}_s^2$ stands for the predicted mean,     as a function of the number of observed
stars $N$ from 3 to 35. The $N=3$ point lies below the training range and is included to assess
generalization to smaller observation sets, as in Fig.~4. For each $N$, 20 independent observation sets were generated for every
test EoS, and the error was averaged over the test set and the realizations.
The thick curves include the complete density grid, whereas the thin curves inside the red-shaded band show the error restricted to the physically
supported region, $n\leq n_{\mathrm{max}}$. Excluding the
extrapolation region
substantially reduces the error; for example, at $N=20$ and
$\sigma_R=0.2\,\mathrm{km}$, it decreases from approximately $0.108 c^2$ to
$0.067 c^2$ for the $R(M)$ input and from $0.098c^2$ to $0.056c^2$ for
$R(M)+\Lambda(M)$.
In both evaluations, the error decreases as a function of $N$, with diminishing improvements as $N$ increases. Including
$\Lambda(M)$ consistently lowers the error at every observation count and
noise level, and the improvement persists even in the noise-free case; its
magnitude is comparable across the noise scenarios rather than growing with
the noise. Thus, tidal deformability provides important additional
information for sound speed reconstruction, although its absolute effect is
smaller than that for the pressure reconstruction.
No saturation is observed within the explored $N$ range.
Although $N=3$ lies below the training range, once again the model still produces
finite and physically meaningful reconstructions, with errors that remain
moderate and decrease as the number of observations increases.

\subsubsection{Density sensitivity to observations}
\label{sec:dens_sens_obs}

\begin{figure*}[!t]
    \centering
    \includegraphics[width=0.8\linewidth]{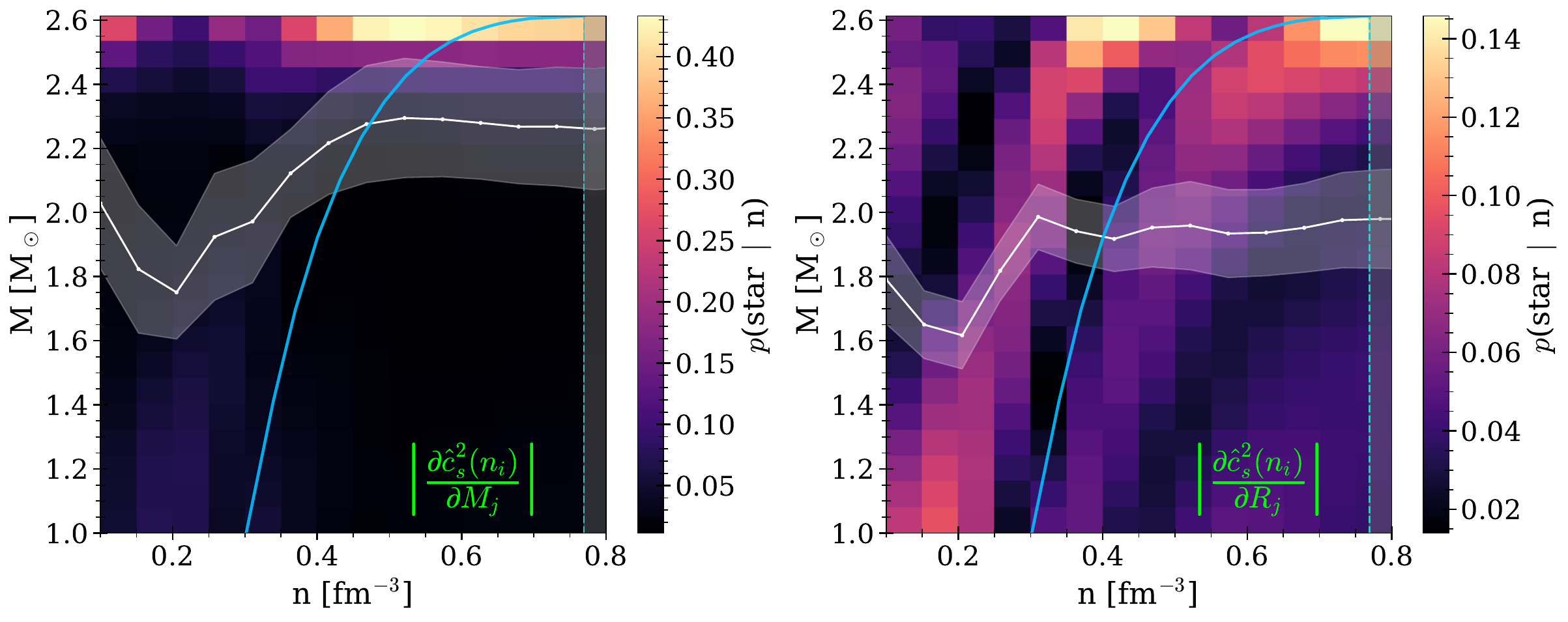}
\caption{Sensitivity of the \texttt{R-$\sigma$} model ($R(M)$ input data,
$\sigma_R = 0.2\,\mathrm{km}$) to the mass and radius of the observed stars for the sound speed squared target, evaluated for a reference GP EoS
with $M_{\mathrm{max}} = 2.61\,M_{\odot}$ and
$n_{\mathrm{max}} = 0.77\,\mathrm{fm}^{-3}$. Each panel shows one input
channel in its own physical units: the mass channel
$|\partial\hat{c}_s^2(n_i)/\partial M_j|$ (left) and the radius channel
$|\partial\hat{c}_s^2(n_i)/\partial R_j|$ (right), with the derivatives
taken with respect to the raw physical observables. The colour encodes
$p(\text{star}\mid n)$, the share of that channel's sensitivity carried by
stars of each mass at density $n$. The sensitivity-weighted mean mass,
i.e.\ the centre of mass of $p(\text{star}\mid n)$, and its $1\sigma$
spread are shown in white, and the EoS's central-density--mass relation
$n_c(M)$ in cyan. The dashed line and gray band mark the extrapolation
region $n > n_{\mathrm{max}}$. See text for details.}
    \label{fig:V4}
\end{figure*}

The analysis of Figs.~\ref{fig:A4} and~\ref{fig:A5} can be repeated for the
sound speed target: the predicted pressure $\hat{P}$ in
Eqs.~\eqref{eq:sensitivity} and~\eqref{eq:sensitivityC} is replaced by the
predicted sound speed squared $\hat{c}_s^2$. Figure~\ref{fig:V4} shows the
two sensitivity channels for the \texttt{R-$\sigma$} model
($\sigma_R=0.2\,\mathrm{km}$), evaluated for a reference GP EoS with
$M_{\mathrm{max}}=2.61\,M_{\odot}$ and $n_{\mathrm{max}}=0.77\,
\mathrm{fm}^{-3}$; left) the mass channel $s^M$, right) the radius channel
$s^R$. There are some clear differences between these results and the pressure ones (see Figs.~\ref{fig:A4} and~\ref{fig:A5}) that can be traced back to the generating properties of the GP EoS ensemble. 
The mass channel again shows a rising band, but here it reflects only a
mass-ordering of the densities: individual stars do not peak at their own
central density. The radius channel carries that association instead:
each star's sensitivity is largest near the star's own central density.

\begin{figure*}[!t]
    \centering
    \includegraphics[width=0.8\linewidth]{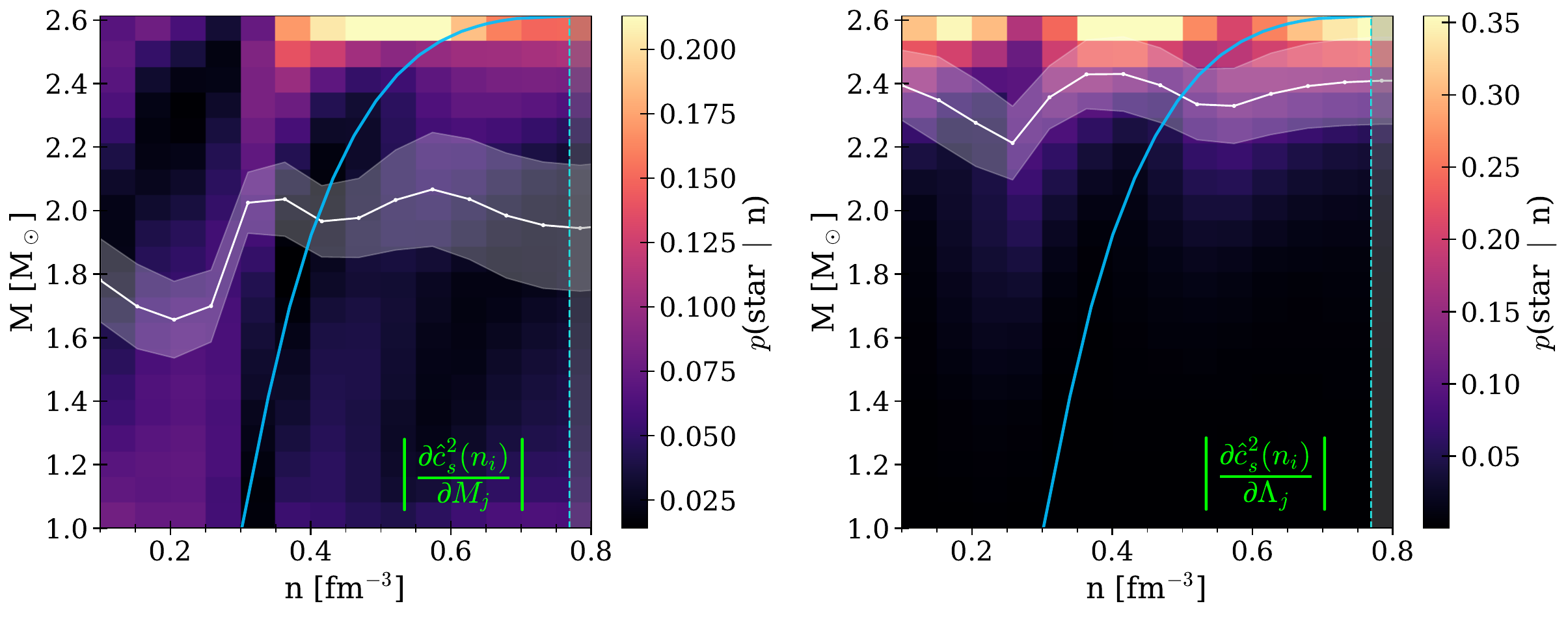}
\caption{Same construction as Fig.~\ref{fig:V4} (same reference EoS, probe
and plotting conventions), now for the \texttt{$\Lambda$-$\sigma$} model,
trained on $\Lambda(M)$ observations with $\sigma_\Lambda = 20\%$ relative
noise instead of the $R(M)$ observations of the \texttt{R-$\sigma$} model.
The right panel shows the tidal-deformability channel
$|\partial\hat{c}_s^2(n_i)/\partial\Lambda_j|$ in place of the radius
channel, the derivative being taken with respect to the physical
$\Lambda$.}
    \label{fig:V5}
\end{figure*}

Figure~\ref{fig:V5} shows the corresponding channels for the
\texttt{$\Lambda$-$\sigma$} model, trained with $\sigma_\Lambda=20\%$
relative noise and evaluated on the same reference EoS. Both panels show a
rising band, but the weakest alignment with the true $n_c(M)$ relation of
the four maps. In the mass channel, the brightest mass bin is the heaviest
stars at essentially every density, and individual stars do not peak at
their own central density: the band reflects that the model leans
increasingly on massive stars as density grows, not that each star probes
the density it reaches. The $\Lambda$ channel is weak and concentrated on
the heaviest stars as well. 

Both speed of sound speed maps also show a marked oscillatory structure, with the
sensitivity varying from one density column to the next in a wavy pattern
absent from the pressure maps. The cause lies in the training data. The
sound speed family is the GP ensemble of \cite{Annala2023},
whose prior follows the construction of \cite{Gorda2023,komoltsev_2023_10101447}: the sound speed 
enters through $\phi(n) = -\ln(1/c_s^2 - 1)$, on which a
squared-exponential kernel acts with a correlation length drawn from
$l \sim \mathcal{N}(1.0\,n_s,\,(0.25\,n_s)^2)$,
$n_s = 0.16\,\mathrm{fm}^{-3}$, around a central value
$\bar{c}_s^2 = 0.5 c^2$. Measured directly on the training set, with no model
involved, the correlation between the sound speed at two densities decays
with a $1/e$ length of $0.22\,\mathrm{fm}^{-3}$, matching the
$\sqrt{2}\,l$ expected for a squared-exponential kernel, and a typical
curve carries two or three turning points. The pressure target, by
contrast, is strictly monotone and contains no such scale. The wavy
features of the sound speed maps are therefore the imprint of the training
dataset: the models reflect the structure they were trained on rather than
introducing it.

\subsubsection{Compactness}

\begin{figure*}[!t]
    \centering
    \includegraphics[width=0.9\linewidth]{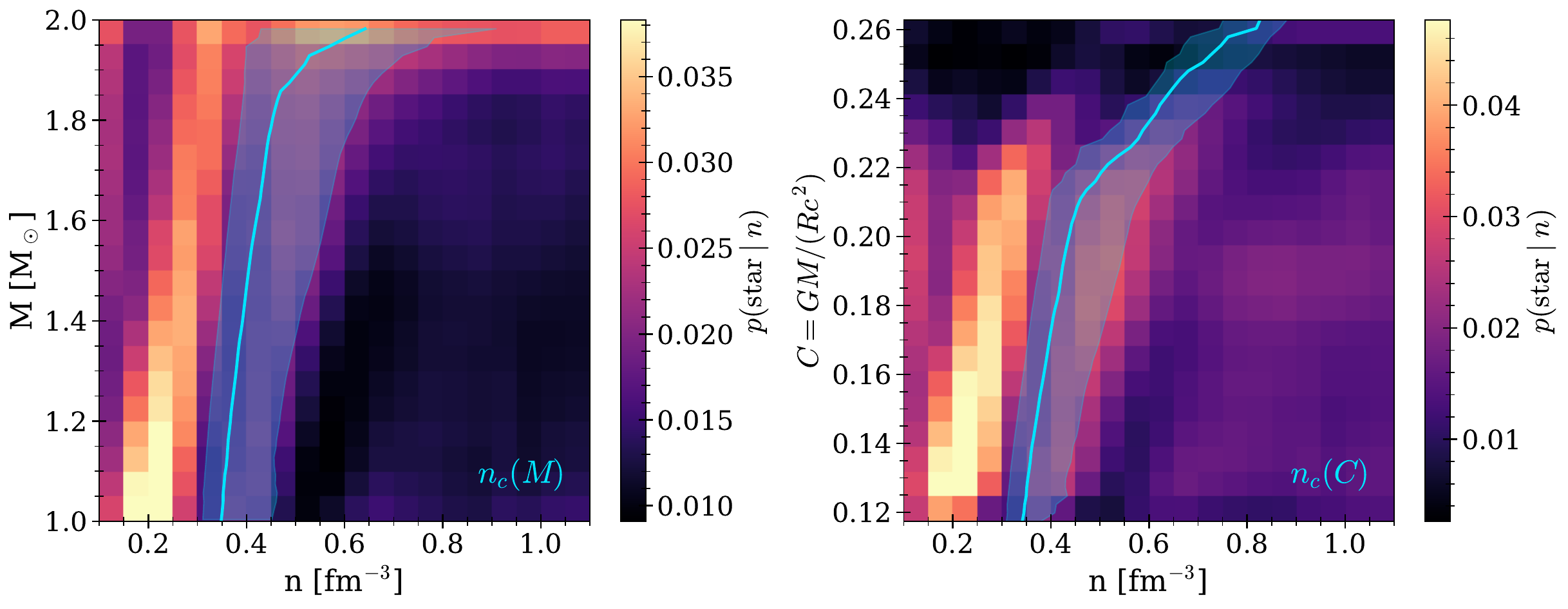}
\caption{Ensemble-averaged sensitivity of the \texttt{R-$\sigma$} model
($R(M)$ input data, $\sigma_R = 0.2\,\mathrm{km}$) for the sound speed
target, with the two channels combined through Eq.~\eqref{eq:sensitivityC}
and averaged over the $7\,798$ GP test EoS. The colour
encodes $p(\text{star}\mid n)$, the share of the combined sensitivity
carried by stars of a given mass (left) or compactness (right) at density
$n$. The cyan curve and band show the median and 10th--90th-percentile
range of the central-density relations $n_c(M)$ and $n_c(C)$ across the
ensemble.}
    \label{fig:V6}
\end{figure*}

The quasi-universality of compactness is not specific to the polytropic
parametrization. Figure~\ref{fig:V6} repeats the analysis of
Fig.~\ref{fig:A6} for the GP ensemble and the sound speed squared
target, with the predicted pressure in Eq.~\eqref{eq:sensitivityC} replaced
by the predicted sound speed $\hat{c}_s^2$. The probe, the per-EoS
normalization, and the ensemble average are identical to
Fig.~\ref{fig:A6}, now over the $7\,798$ test EoS of the GP
set.

The result parallels the pressure case. Grouping stars by compactness
instead of mass reduces the across-EoS spread of the true central density
from $0.083$ to $0.061\,\mathrm{fm}^{-3}$, a ratio of $0.74$. As in Fig.~\ref{fig:A6}, this model-independent contraction is the decisive
quantity of the figure, and the model map serves only to illustrate it: in
mass coordinates, the band runs loosely within the family of individual
$n_c(M)$ curves, which differ from one EoS to another. While re-binned by
compactness, it follows the universal $n_c(C)$ relation, though more loosely
than in the pressure case.

Finally, as in the single-EoS sound speed maps of Figs.~\ref{fig:V4}
and~\ref{fig:V5}, the ensemble band is not perfectly smooth: its
sensitivity-weighted mean compactness oscillates with density. This
waviness is the imprint of the GP training set, which encodes
the correlation length of about $0.22\,\mathrm{fm}^{-3}$ established above:
the oscillation occurs on exactly that scale and is stable across random
subsets of the ensemble, so it reflects the statistics of the dataset
rather than a property of individual EoS or of the network.

\section{Conclusions}
\label{sec:conclusions}
%======================================================================
We have trained a permutation-invariant Set Transformer to reconstruct NS
EoS from variable-size, unordered sets of stellar observations. The model
accepts observations of $M$ and $R$, $M$ and $\Lambda$, or both, and
predicts the pressure profile $P(n)$ or the squared sound speed profile
$c_s^2(n)$ on a fixed density grid. A heteroscedastic Gaussian output head
provides a density-dependent predictive uncertainty. 
Twenty models corresponding to ten model configurations were trained on two independent EoS ensembles, a
piecewise-polytropic ensemble for the pressure target and a
Gaussian-process (GP) ensemble for the sound speed target,
nine heteroscedastic configurations spanning three noise scenarios and three input combinations, plus the deterministic R-MSE baseline, giving ten configurations per target.

The uncertainty head does not degrade the reconstruction accuracy, while
its predictive intervals are well calibrated for both targets: the nominal
$2\sigma$ intervals contain the true profile at approximately
$95$--$97\%$ of the density points. The predicted uncertainty
also tracks the actual reconstruction error and increases in regions where
the EoS is not constrained by stable stars. 
Adding tidal deformability generally improves the reconstruction at a fixed observation count, including for the squared speed of sound and in the noise-free case, and this improvement persists even when the \(\Lambda\) observations carry their own relative observational noise. In
all configurations, the error decreases as more stars are added, with
progressively smaller gains at larger set sizes. No saturation is observed
within the explored range of the number of observations ($N=5$--$35$), and
the additional information from $\Lambda$ can be comparable to or greater
than that obtained by substantially increasing the number of observations
(i.e. a set with N=10 radius and N=10 tidal-deformability measurements can perform comparably to a set with N=20 radius measurements alone.)

In the pressure models, the sensitivity of each star peaks close to the
central density of that same star.
As a result, the prediction at density $n$ is dominated by the stars that actually reach
$n$, and the informative mass of the mass channel saturates among the
heaviest stable stars, just below $M_{\mathrm{max}}$.
In the sound speed models, the mass channels only order stars by density:
the density at which the sensitivity of a star peaks is essentially
unrelated to the central density of that star. The
radius channel retains this association. The sensitivity of each star
peaks near the central density of the star itself, and the peak
mass rises with density along the expected central-density relation. The
$\Lambda$ channels are instead concentrated on the most massive stars at
all densities: the value of $\Lambda$ itself falls by about three orders
of magnitude along the mass--radius curve, so per unit of $\Lambda$
massive stars carry by far the largest response. The $\Lambda$ channels
saturate at the heaviest stars much as the mass channel does, in part a
consequence of expressing the sensitivity per unit of $\Lambda$. The noisy
$M$--$\Lambda$ models show the same qualitative structure, with their
sensitivity concentrated more strongly toward massive, compact stars.
Thus, the mass channel orders stars by density and relies most heavily on
the heaviest ones, while the radius and tidal-deformability channels
combine these star-by-star peaks with a contribution from stars of all
masses that persists at every density.

Grouping stars by compactness makes this structure nearly
EoS-independent. The across-EoS spread of the true central density at fixed
mass drops by about $30\%$ for the polytropic ensemble
($0.090 \to 0.064\,\mathrm{fm}^{-3}$) and by roughly one quarter for the
GP ensemble ($0.083 \to 0.061\,\mathrm{fm}^{-3}$) when stars are
parametrized by $C = GM/(Rc^2)$ instead of mass. Accordingly, the
ensemble-averaged sensitivity maps, built from the per-star log-sensitivity
of Eq.~\eqref{eq:sensitivityC}, follow a quasi-universal band in
compactness coordinates that tracks the universal $n_c(C)$ relation.

Finally, the sound speed maps inherit the structure of their training
family. The GP ensemble encodes a characteristic correlation length of
$0.22\,\mathrm{fm}^{-3}$ in $c_s^2(n)$, and the population sensitivity maps
oscillate on precisely that scale, stably across random subsets of the
ensemble: the models reproduce the waviness of the training data rather
than introducing structure of their own.

Overall, the sensitivity framework converts the trained network into an
interpretable map showing which star constrains which density. Compactness
emerges as the natural coordinate of this map, while the predicted
uncertainty marks the densities that no stable star can reach. The attention-based design is what gives the model its capacity: all
pairwise nonlinear correlations among the $N$ observations are available to
the network, and each output density reads the full encoded set through its
own learnable query, so the assignment of stars to densities is learned,
not assumed. The sensitivity maps of the previous sections show what this
capacity was used for: the network discovered on its own that a star
constrains the EoS at its own central density, that the heaviest stars
dominate the highest densities, and that this assignment adapts to the
target family and to the observables provided.

\section*{Acknowledgements} 
V.C. expresses sincere gratitude to the FCT for their generous support through Ph.D. grant number 2024.00311.BD. This work was partially supported by national funds from FCT (Fundação para a Ciência e a Tecnologia, I.P, Portugal) under project UID/04564/2025, identified by DOI 10.54499/UIDB/04564/2025, and project  2024.16290.PEX identified by DOI  identifier 10.54499/2024.16290.PEX, by the European Union-Next Generation EU, Mission 4 Component 1 CUP J53D23001550006 with the PRIN Project No. 202275HT58, and the Polish National Science Center OPUS grant no. 2021/43/B/ST9/01714.

\appendix
\section{{Model Calibration}}
\subsection{Pressure}
\begin{figure*}[!ht]
    \centering
    \includegraphics[width=0.80\linewidth]{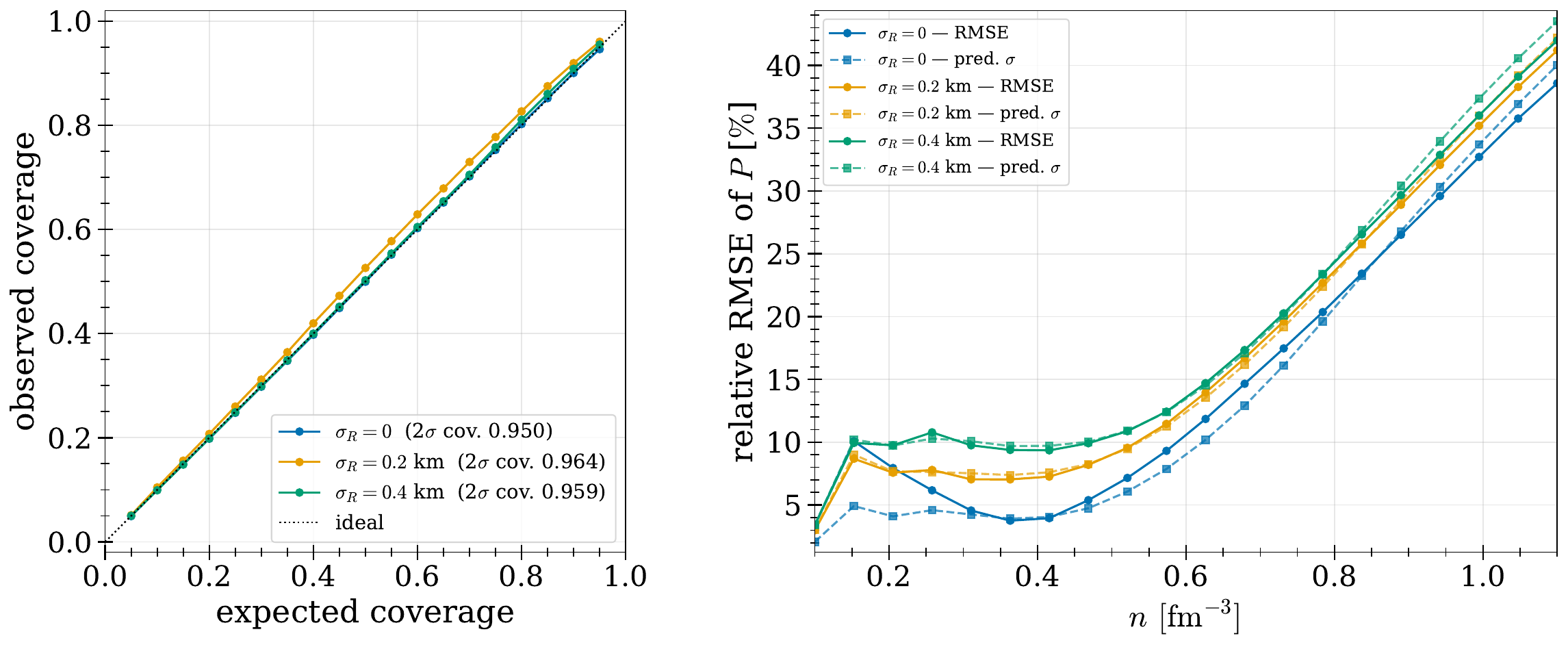}
\caption{Reliability of the predicted uncertainties for the three $R(M)$
    models trained with $\sigma_R=0$ (blue), $0.2$~km (orange), and $0.4$~km
    (green), using one simulated observation set per EoS, with
$N$ drawn uniformly from $\mathcal{U}\{5,35\}$.
    (Left) Calibration curve: observed vs.\ expected coverage of the predicted
    central intervals at 19 confidence levels from $0.05$ to $0.95$, pooled
    over the $4\,030$ polytropic test EoS and their 20 density points, with the $2\sigma$ coverages quoted in the legend
    and the Gaussian expectation $0.9545$.
    (Right) Relative RMSE of the pressure,
    $\big\langle (\hat P/P-1)^2\big\rangle^{1/2}\times100$
    (solid) and
    the mean predicted relative uncertainty (dashed), in percent, as
    functions of density; $\hat{\sigma}$ is the model's predicted standard
    deviation of $\log_{10}P$, converted to a relative error by
    $10^{\hat{\sigma}}-1$ transformation, and both curves are averaged over
    the test EoS at each density.}
    \label{fig:A1}
\end{figure*}

Figure~\ref{fig:A1} assesses the reliability of the predicted uncertainties
for three heteroscedastic $M$--$R$ models that differ only in the radius
noise used during training: \texttt{R-0} ($\sigma_R=0$, blue),
\texttt{R-$\sigma$} ($\sigma_R=0.2\,\mathrm{km}$, orange), and
\texttt{R-$2\sigma$} ($\sigma_R=0.4\,\mathrm{km}$, green). In the left
panel, the nominal $2\sigma$ interval covers the true pressure at
$95.0\%$, $96.4\%$, and $95.9\%$ of test-set density points, respectively,
close to the Gaussian expectation of $95.4\%$; the mean calibration errors
are $0.002$, $0.018$, and $0.004$, indicating good overall calibration.

The right panel compares, at each density, the relative RMSE of the pressure over the test EoS (solid curves) with the mean
predicted relative uncertainty, $(10^{\hat \sigma}-1)\times100$ (dashed
curves). The two noise-trained models track the error closely across the
density grid. The noise-free model underestimates the error near
$n \simeq 0.15$--$0.2\,\mathrm{fm}^{-3}$, where the error is approximately
twice its predicted uncertainty. For all models, both quantities increase
toward high densities, reaching roughly $40\%$, as fewer observations
constrain the highest-density region. Thus, training with realistic radius
noise produces uncertainty estimates that better reflect the actual
reconstruction error.

\begin{figure*}[!ht]
    \centering
\includegraphics[width=0.50\linewidth]{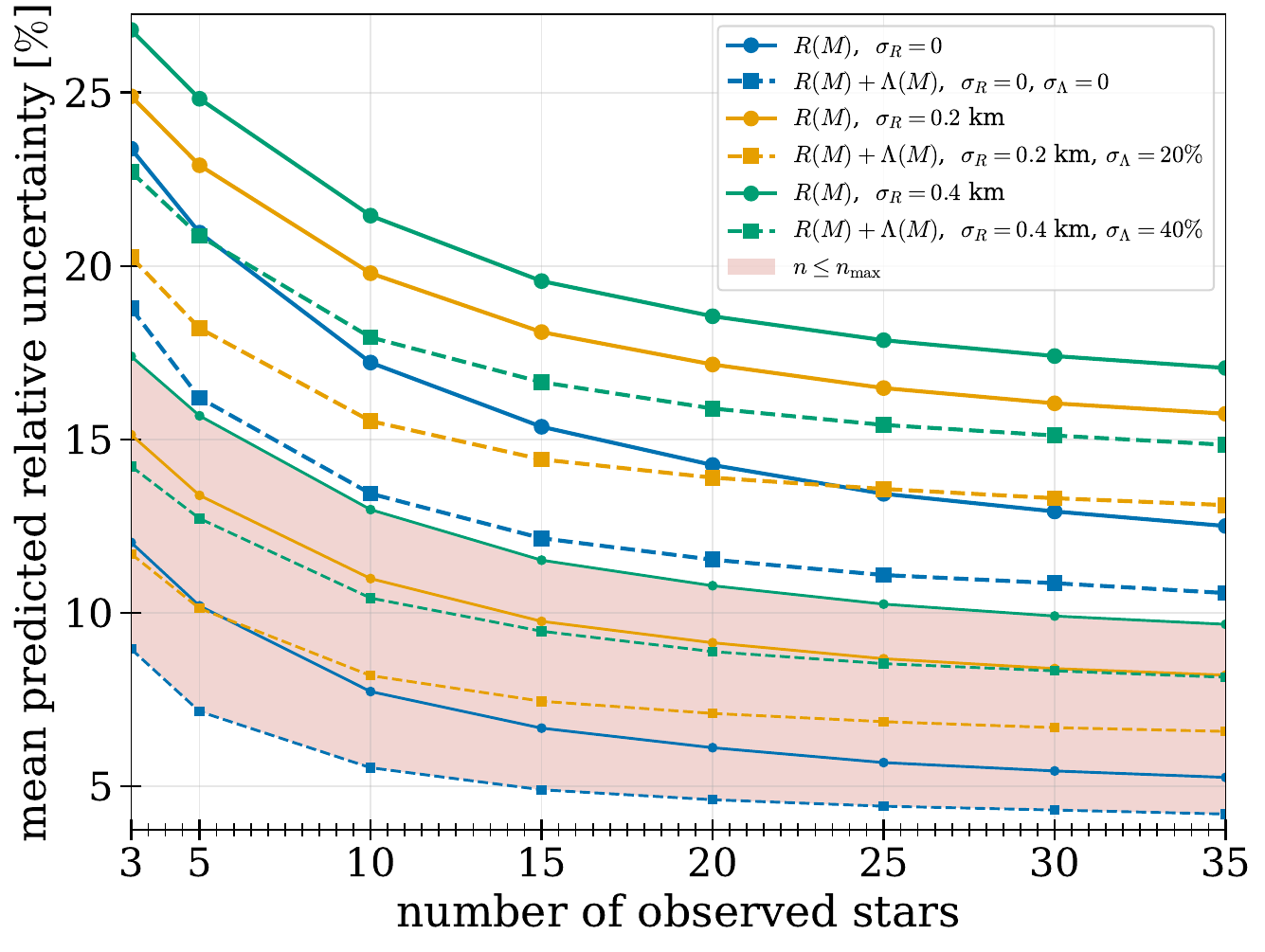}
\caption{Mean predicted relative pressure uncertainty,
    $\langle 10^{\hat{\sigma}}-1\rangle\times100$ with $\hat{\sigma}$ the
    model's predicted standard deviation of $\log_{10}P$, over the test set as a
    function of the number of observations.  Colors denote the three radius-noise scenarios:
\texttt{R-0} ($\sigma_R=0$, blue), \texttt{R-$\sigma$}
($\sigma_R=0.2\,\mathrm{km}$, orange), and \texttt{R-$2\sigma$}
($\sigma_R=0.4\,\mathrm{km}$, green). Solid lines with circles correspond to
$R(M)$, while dashed lines with squares correspond to $R(M)+\Lambda(M)$. Thick curves use the complete
density grid, including the extrapolation region; thin curves of the same line style and the
red-shaded band are restricted to the physically supported region,
$n\leq n_{\mathrm{max}}$.  }
    \label{fig:A3_robustness_unc}
\end{figure*}

Figure~\ref{fig:A3_robustness_unc} shows the mean predicted relative pressure
uncertainty as a function of the number of observations. For each of the
three radius-noise scenarios and both input {modes}, we generated
one observation set for each test EoS at each set size
{$N = 3, 5, 10, \dots, 35$}, as in Fig. \ref{fig:A3}, the \(N=3\) point lies below the training range and is included to assess generalization to smaller observation sets. If $\hat{\sigma}(n)$ denotes the
predicted uncertainty in $\log_{10}P$ at density $n$, we converted it to a
fractional pressure uncertainty according to
$u_P(n)=\left(10^{\hat{\sigma}(n)}-1\right)\times100\%$.
The plotted quantity is the average of $u_P(n)$ over all test EoS and density
points. Thick curves include the complete density grid, including the
extrapolation region. Thin curves inside the red-shaded band show the same
average restricted to densities supported by the corresponding EoS,
$n\leq n_{\mathrm{max}}$. Thus, the figure measures the uncertainty reported
by the model, rather than the reconstruction error itself, and shows how this
predicted uncertainty changes as more stars are observed.
The predicted uncertainty decreases as more stars are observed, with the
$R(M)+\Lambda(M)$ models generally reporting smaller uncertainties and the supported-density estimates remaining below the
full-grid values. The out-of-training-range case $N=3$ also gives reasonable results,
indicating generalization to smaller observation sets.

\subsection{Sound speed}
\begin{figure*}[!t]
    \centering
    \includegraphics[width=0.7\linewidth]{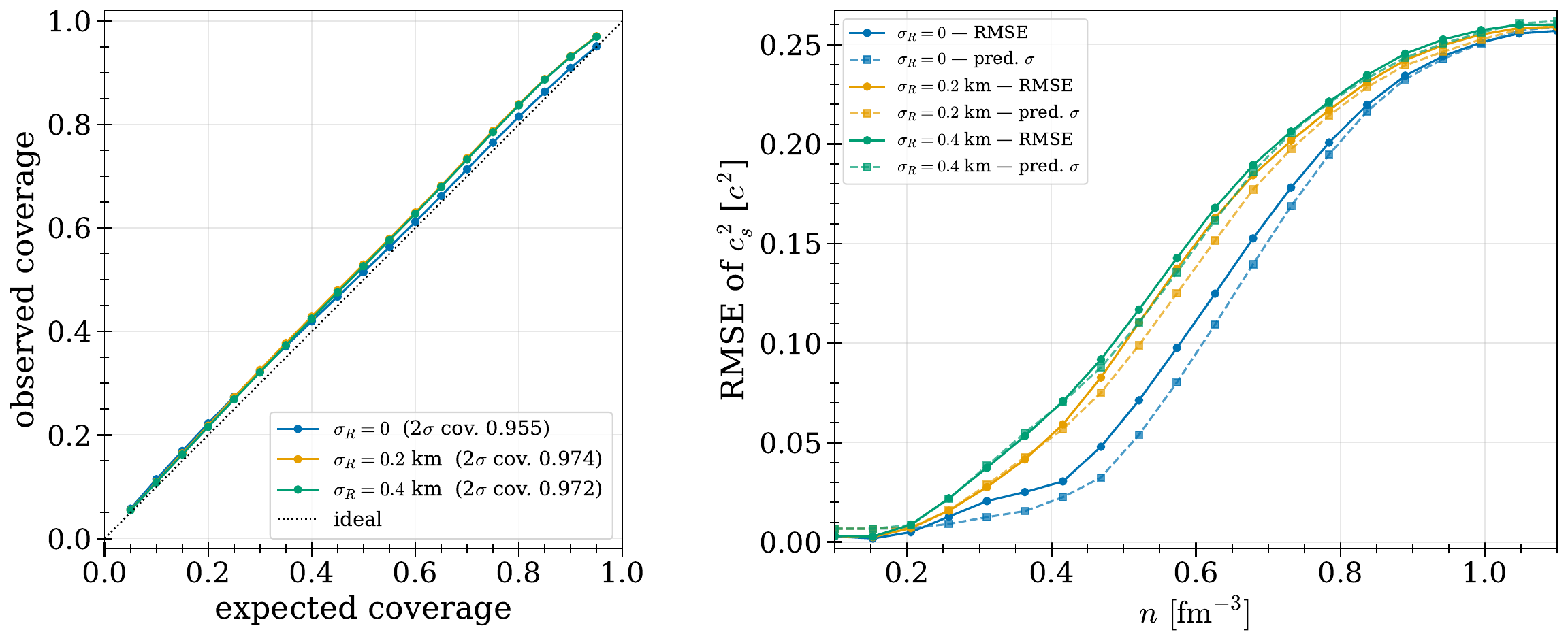}
\caption{Reliability of the predicted uncertainties for the three $R(M)$
    models trained on the GP ensemble with $\sigma_R=0$ (blue), $0.2$~km
    (orange), and $0.4$~km (green), using one simulated set per EoS, with
$N$ drawn uniformly from $\mathcal{U}\{5,35\}$.
    (Left panel) Reliability diagram: observed vs.\ expected coverage of the
    predicted central intervals at 19 confidence levels from $0.05$ to $0.95$,
    pooled over the $7\,798$ GP test EoS and their 20 density points, with the $2\sigma$ coverages quoted in the legend
    and the Gaussian expectation $0.9545$.
    (Right panel) RMSE of $c_s^2$,
    $\big\langle (\hat{c}_s^2-c_{s}^2)^2\big\rangle^{1/2}$ (solid lines), and
    the mean predicted uncertainty $\langle\hat{\sigma}\rangle$ (dashed lines),
    both in units of $c^2$ and averaged over the test EoS at each density, as
    functions of density $n$; $\hat{\sigma}$ is the model's predicted standard
    deviation of $c_s^2$ and, unlike the pressure panel Fig.~\ref{fig:A1}, no logarithmic
    transformation is involved.}
    \label{fig:V1}
\end{figure*}

Figure~\ref{fig:V1} assesses the reliability of the predicted uncertainties
for three heteroscedastic $R(M)$ models trained to reconstruct the
sound speed profile. They differ only in the radius noise used during
training: \texttt{R-0} ($\sigma_R=0$, blue), \texttt{R-$\sigma$}
($\sigma_R=0.2\,\mathrm{km}$, orange), and \texttt{R-$2\sigma$}
($\sigma_R=0.4\,\mathrm{km}$, green). The nominal $2\sigma$ interval
contains the true $c_s^2$ at $95.5\%$, $97.4\%$, and $97.2\%$ of test-set
density points, respectively, compared with the Gaussian expectation of
$95.4\%$. The corresponding mean calibration errors are $0.015$, $0.026$,
and $0.024$. Thus, all three models are reasonably calibrated, although
the noise-trained models slightly overestimate their uncertainty.

The right panel shows the density dependence of the achieved error and
predicted uncertainty, both in units of $c^2$. Both are very small at low
densities and increase toward the upper end of the grid, where fewer
observations probe the relevant central densities. The predicted
uncertainty follows the achieved error reasonably well for the two
noise-trained models, while \texttt{R-0} underestimates the error over
part of the intermediate-density range, $n\simeq0.26$--$0.52\ \fmc$,
where the error reaches up to $1.6$ times its predicted uncertainty. The
heteroscedastic output, therefore, provides a useful density-resolved
uncertainty estimate for the bounded sound speed squared target.

\begin{figure*}[!ht]
    \centering
\includegraphics[width=0.50\linewidth]{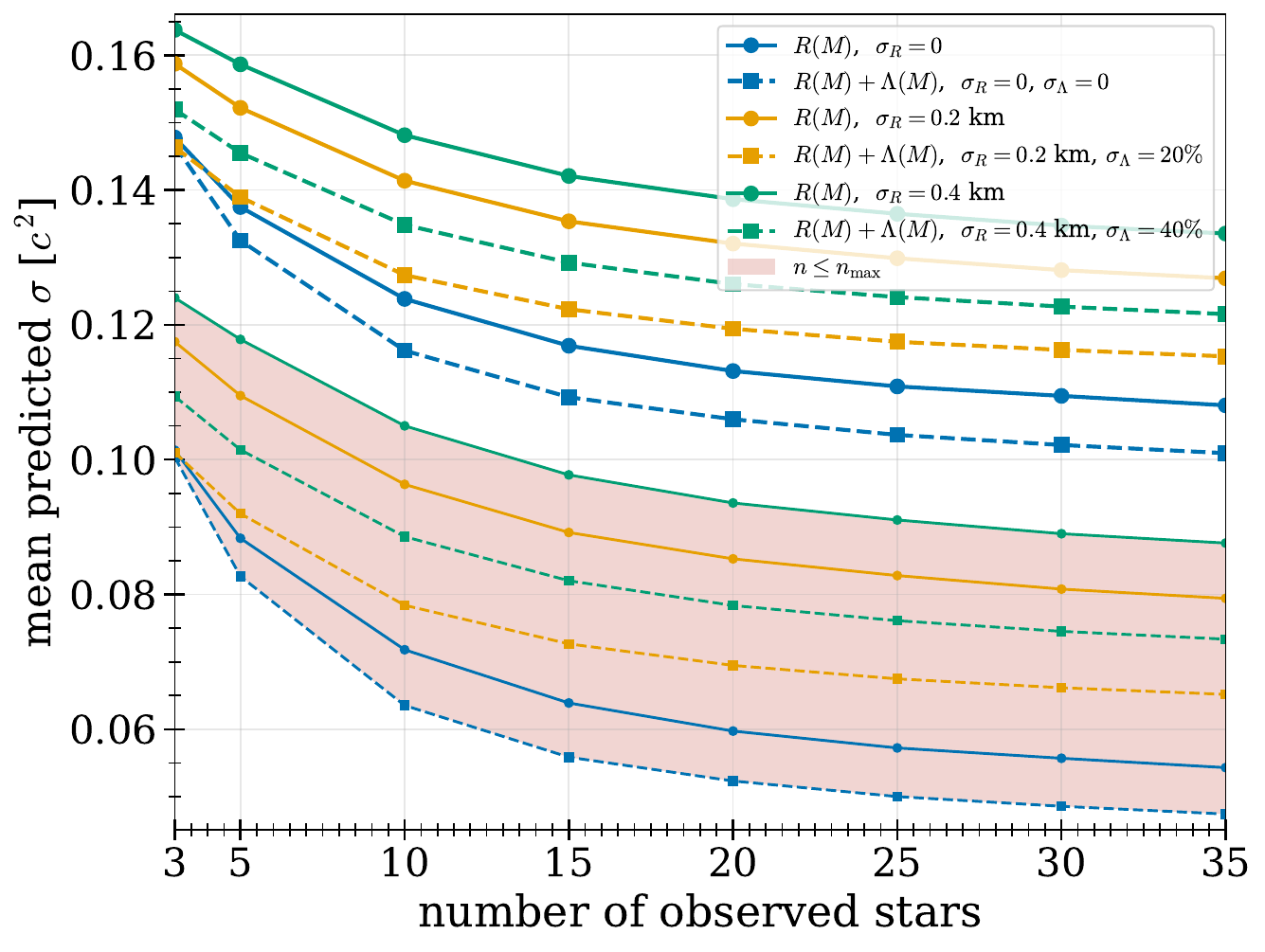}
    \caption{Mean predicted uncertainty of $c_s^2$,
    $\langle\hat{\sigma}\rangle$ in units of $c^2$ with $\hat{\sigma}$ the
    model's predicted standard deviation of $c_s^2$ (no logarithmic
    transformation is involved), over the test set as a function of the number
    of observations. Colors denote the three radius-noise scenarios:
    \texttt{R-0} ($\sigma_R=0$, blue), \texttt{R-$\sigma$}
    ($\sigma_R=0.2\,\mathrm{km}$, orange), and \texttt{R-$2\sigma$}
    ($\sigma_R=0.4\,\mathrm{km}$, green). Solid lines with circles correspond to
$R(M)$, while dashed lines with squares correspond to $R(M)+\Lambda(M)$. Thick curves use the complete
density grid, including the extrapolation region; thin curves of the same line style and the
red-shaded band are restricted to the physically supported region,
$n\leq n_{\mathrm{max}}$.} \label{fig:V3_robustness_unc}
\end{figure*}

Figure~\ref{fig:V3_robustness_unc} shows the mean predicted uncertainty of
$c_s^2$ as a function of the number of observations. The curves correspond
to the three radius-noise scenarios, \texttt{R-0}, \texttt{R-$\sigma$},
and \texttt{R-$2\sigma$}, and to the two input modes, $R(M)$ and
$R(M)+\Lambda(M)$ (for the combined mode, $N$ counts the observations of
each type, as in Fig.~\ref{fig:A3}). For each set size
$N = 3, 5, 10, \dots, 35$, the model-predicted standard deviation was
averaged over the test EoS and density points, in units of $c^2$.
Thick curves cover the complete density grid, including the extrapolation
region. Thin curves inside the red-shaded band restrict the average to
densities supported by the corresponding EoS, $n\leq n_{\mathrm{max}}$.
The predicted uncertainty decreases as more stars are observed and is
smaller for the $R(M)+\Lambda(M)$ models at all set sizes, while the
supported-density estimates remain below the full-grid values.
For $N=3$, the model again gives reasonable results, indicating
generalization to a set size below the training range.

\bibliography{biblio}

\end{document}